\documentclass{svjour3}                     % onecolumn (standard format)

\usepackage{graphicx}
\usepackage{natbib}
\bibpunct{(}{)}{;}{a}{}{,}

\newcommand{\be}{\begin{equation}}
\newcommand{\ee}{\end{equation}}
\newcommand{\beq}{\begin{eqnarray}}
\newcommand{\eeq}{\end{eqnarray}}

\newcommand{\dd}{\mathrm{d}}

\newcommand{\ion}[2]{#1\,{\sc{#2}}}

\renewcommand{\vec}[1]{\bf{#1}}

\journalname{SSRv}

\begin{document}

\title{Simulation techniques for cosmological simulations}

\author{K.~Dolag,
        S.~Borgani,
        S.~Schindler,
        A.~Diaferio,
        A.M.~Bykov}

\institute{
K. Dolag
\at Max-Planck-Institut f\"ur Astrophysik,
P.O. Box 1317, D-85741 Garching, Germany\\
\email{kdolag@mpa-garching.mpg.de}
\and
S. Borgani
\at
Department of Astronomy, University of Trieste, via Tiepolo 11,
I-34143 Trieste, Italy\\
\email{borgani@oats.inaf.it}
\and
S. Schindler
\at
Institut f\"ur Astro- und Teilchenphysik, Universit\"at Innsbruck,
Technikerstr. 25, 6020 Innsbruck, Austria\\
\email{sabine.schindler@uibk.ac.at}
\and
A. Diaferio
\at Dipartimento di Fisica Generale ``Amedeo Avogadro'', Universit\`a
degli Studi di Torino, Torino, Italy \\
Istituto Nazionale di Fisica Nucleare (INFN), Sezione di Torino, Via
P. Giuria 1, I-10125, Torino, Italy\\
\email{diaferio@ph.unito.it}
\and
A.M. Bykov
\at A.F. Ioffe Institute of Physics and Technology, St. Petersburg,
194021, Russia\\
\email{byk@astro.ioffe.ru}
}

\date{Received: 13 November 2007; Accepted: 14 December 2007 }

\maketitle

\begin{abstract}
Modern cosmological observations allow us to study in great detail the
evolution and history of the large scale structure hierarchy. The
fundamental problem of accurate constraints on the cosmological
parameters, within a given cosmological model, requires precise
modelling of the observed structure. In this paper we briefly
review the current most effective techniques of large scale
structure simulations, emphasising both their advantages and
shortcomings. Starting with basics of the direct N-body simulations
appropriate to modelling cold dark matter evolution, we then discuss
the direct-sum technique {\sl GRAPE}, particle-mesh ({\sl PM}) and hybrid
methods, combining the {\sl PM} and the tree algorithms. Simulations of
baryonic matter in the Universe often use hydrodynamic codes based
on both particle methods that discretise mass, and
grid-based methods. We briefly describe Eulerian grid methods,
and also some variants of Lagrangian smoothed particle hydrodynamics
({\sl SPH}) methods.

\keywords{cosmology: theory \and large-scale structure of universe \and
hydrodynamics \and  method: numerical, N-body simulations}
\end{abstract}

\section{Introduction}

In the hierarchical picture of structure formation, small objects
collapse first and then merge to form larger
and larger structures in a complex manner. This formation process
reflects on the intricate structure of galaxy clusters, whose properties depend on
how the thousands of smaller
objects that the cluster accretes are destroyed or survive
within the cluster gravitational potential.
These merging events are the source of
shocks, turbulence and acceleration of relativistic particles in
the intracluster medium, which, in turn, lead to a redistribution or
amplification of magnetic fields, and to the acceleration of cosmic rays.
In order to model these processes realistically, we need to resort
to numerical simulations which are capable of resolving and
following correctly the highly non-linear dynamics.
In this paper, we briefly describe
the methods which are commonly used to simulate
galaxy clusters within a cosmological context.

Usually, choosing the simulation setup is a compromise between the size of the
region that one has to simulate to fairly represent the object(s) of interest,
and the resolution needed to resolve the objects at the required level of
detail. Typical sizes of the simulated volume are a megaparsec scale for an
individual galaxy, tens to hundreds of megaparsecs for a galaxy population, and
several hundreds of megaparsecs for a galaxy cluster population. The mass
resolution varies from $\approx 10^5$~M$_\odot$ up to $\approx
10^{10}$~M$_\odot$, depending on the object studied, while, nowadays, one can
typically reach the resolution of a few hundred parsec for individual galaxies
and above the kiloparsec scale for cosmological boxes.

\section{N-Body (pure gravity)}

Over most of the cosmic time of interest for structure formation, the
Universe is dominated by dark matter. The most favourable model
turned out to be the so-called cold dark matter (CDM) model. The CDM can be
described as a collisionless, non-relativistic fluid of particles of mass $m$,
position ${\vec x}$ and momentum ${\vec p}$.
In an expanding background Universe (usually described by a
Friedmann-Lema{\^i}tre model), with $a=(1+z)^{-1}$ being the Universe scale factor,
${\vec x}$ is the comoving position and the phase-space distribution
function $f({\vec x},{\vec p},t)$ of the dark-matter fluid can be described by the
collisionless Boltzmann (or Vlasov) equation
\begin{equation}
   \frac{\partial f}{\partial t} + \frac{{\vec p}}{m a^2} {\vec \nabla} f
 - m {\vec \nabla}\Phi \frac{\partial f}{\partial {\vec p}} = 0
\end{equation}
coupled with the Poisson equation
\begin{equation}
   {\vec \nabla}^2 \Phi({\vec x},t) = 4\pi {\rm G}
   a^2\left[\rho({\vec x},t)-\bar{\rho}(t)\right],
\end{equation}
where $\Phi$ is the gravitational potential and $\bar{\rho}(t)$ is the background density.
The proper mass density
\begin{equation}
   \rho({\vec x},t) = \int{f({\vec x},{\vec p},t)
   \dd^3p}
\end{equation}
can be inferred by integrating the distribution function over the
momenta ${\vec p} = m a^2\dot{{\vec x}}$.

This set of equations represents a high-dimensional problem. It is therefore usually solved by
sampling the phase-space density by a finite number $N$ of tracer
particles. The solution can be found through the equation of
motion of the particles (in comoving coordinates),
\begin{equation}
   \frac{\dd{\vec p}}{\dd t} = -
   m {\vec \nabla}\Phi
\end{equation}
and
\begin{equation}
   \frac{\dd{\vec x}}{\dd t} =
   \frac{{\vec p}}{m a^2}.
\end{equation}
Introducing the proper peculiar velocity ${\vec v} = a
\dot{{\vec x}}$ these equations can be written as
\begin{equation}
  \frac{\dd{\vec v}}{\dd t} + {\vec v}\frac{\dot{a}}{a}
  = - \frac{{\vec \nabla}\Phi}{a}.
\end{equation}
The time derivative of the expansion parameter, $\dot{a}$, can be
obtained from the Friedmann equation
\begin{equation}
   \dot{a} = H_0\sqrt{1+\Omega_0(a^{-1}-1)+\Omega_\Lambda(a^2-1)},
\end{equation}
where we have assumed the dark energy to be equivalent to a
cosmological constant. For a more detailed description of the
underlying cosmology and related issues, see for example \citet{
1980lssu.book.....P} or others.

There are different approaches: to solve directly the motion of the
tracer particles, or to solve the Poisson equation. Some of the
most common methods will be described briefly in the following
sections.

\subsection{Direct sum (GRAPE, GPU)}

The most direct way to solve the N-body problem is to sum
directly the contributions of all the individual particles to the
gravitational potential
\begin{equation}
\Phi({\vec r})=-{\rm G}\sum_j\frac{m_j}{\left(|{\vec r}-{\vec r}_j|^2 +
\epsilon^2\right)^{\frac{1}{2}}}. \label{eq_plum}
\end{equation}
In principle, this sum would represent the exact (Newtonian) potential
which generates the
particles' acceleration. As mentioned before, the particles do not
represent individual dark matter particles, but should be
considered as Monte Carlo realisations of the mass distribution, and
therefore only collective, statistical properties can be considered.
In such simulations, close encounters between individual
particles are irrelevant to the physical problem under
consideration, and the gravitational force between two particles is
smoothed by introducing the gravitational softening $\epsilon$. This
softening
reduces the spurious two-body relaxation which occurs when the number
of particles in the simulation is not large enough to represent
correctly a collisionless fluid. This situation however is unavoidable,
because the number of dark matter particles in real systems is
orders of magnitude larger than the number that can be handled in
a numerical simulation.
Typically, $\epsilon$ is chosen to
be $1/20-1/50$ of the mean inter-particle separation within the
simulation. In general, this direct-sum approach is considered
to be the most accurate technique, and is used for problems where
superior precision is needed. However this method has the disadvantage of
being already quite CPU intensive for even a moderate number of particles,
because the computing time is $\propto N^2$, where $N$ is the total number
of particles.

Rather than searching for other software solutions,
an alternative approach to solve the
$N^2$-bottleneck of the direct-sum technique
is the {\sl GRAPE} (GRAvity PipE) special-purpose hardware (see
e.g. \citealt{1993PASJ...45..339I} and related articles). This hardware is
based on custom chips that compute the gravitational force
with a hardwired Plummer force law (Eq.~\ref{eq_plum}).
This hardware device thus solves
the gravitational N-body problem with a direct summation
approach at a computational speed which is considerably higher than
that of traditional processors.

For the force computation, the particle coordinates are first
loaded onto the {\sl GRAPE} board, then the forces for several positions
(depending on the number of individual {\sl GRAPE} chips installed in
the system) are computed in parallel. In practice, there are some
technical complications when using the {\sl GRAPE} system. One is that
the hardware works internally with special fixed-point formats or
with limited floating point precision (depending on the version of
the {\sl GRAPE} chips used) for positions, accelerations and masses.
This results in a reduced dynamic range compared to the standard IEEE
floating point arithmetic. Furthermore, the communication time
between the host computer and the {\sl GRAPE} system can be an issue in
certain circumstances. However, newer versions of the {\sl GRAPE} chips
circumvent this problem, and can also be combined
with the tree algorithms (which are described in detail in the next
section), see
\citet{1991PASJ...43..841F,1991PASJ...43..621M,1998MNRAS.293..369A,2000PASJ...52..659K}.

By contrast, the graphic processing unit ({\sl GPU}) on modern graphic
cards now provides an alternative tool for high-performance computing.
The original purpose of the {\sl GPU} is to serve as a graphics accelerator
for speeding up the image processing, thereby allowing one to perform
simple instructions on multiple data. It has therefore become an
active area of research to use the {\sl GPU}s of the individual members
of computer clusters. Although very specialised, many of those computational
algorithms are also needed in computational astrophysics, and therefore
the {\sl GPU} can provide
significantly more computing power than the host system;
thereby providing a high performance with typically large memory
size and at relatively low cost, which represents a valid alternative to special purpose
hardware like {\sl GRAPE}. For recent applications to astrophysical
problems see \citet{2007arXiv0707.2991S} and references therein.

\subsection{Tree}
\begin{figure*}
\rotatebox{90}{\resizebox{4cm}{!}{\includegraphics{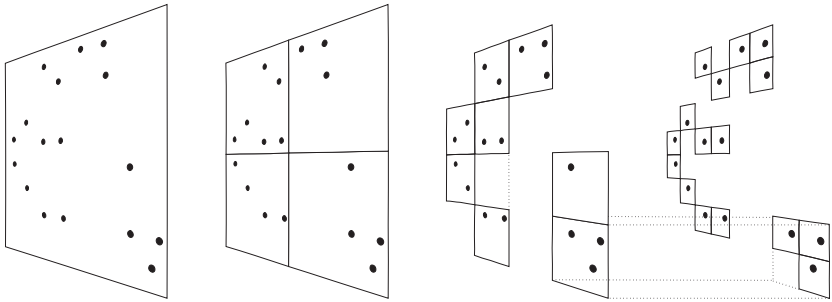}}}\\
\caption{Schematic illustration of the \protect\citet{1986Natur.324..446B}
oct-tree in
two dimensions.  The particles are first enclosed in a square
(root node). This square is then iteratively subdivided into four
squares of half the size, until exactly one particle is left in
each final square (leaves of the tree). In the resulting tree
structure, each square can be the progenitor of up to four siblings.
Taken from \protect\citet{springel01}.} \label{tree_construct}
\end{figure*}

The primary method of solving the N-body problem is a hierarchical
multipole expansion, commonly called a tree algorithm. This method
groups distant particles into larger cells, allowing their gravity
to be accounted for by means of a single multipole force. Instead
of requiring $N-1$ partial force evaluations per particle, as needed in a
direct-summation approach, the gravitational force on a single
particle can be computed with substantially fewer operations, because distant
groups are treated as ``macro'' particles in the sum. In this manner the
sum usually reduces to $N\mathrm{log}(N)$ operations. Note
however that this scaling is only true for homogeneous particle distributions, 
whereas the scaling for strongly inhomogeneous distributions, as
present in evolved cosmological structures, can be less efficient.

In practice, the hierarchical grouping that forms the basis of the
multipole expansion is most commonly obtained by a recursive
subdivision of space. In the approach of
\citet{1986Natur.324..446B}, a cubical root node is used to
encompass the full mass distribution; the cube is repeatedly
subdivided into eight daughter nodes of half the side-length each,
until one ends up with `leaf' nodes containing single particles
(see Fig.~\ref{tree_construct}).
Forces are then obtained by ``walking'' the tree. In other
words, starting at
the root node, a decision is made as to whether or not the multipole
expansion of the node provides an accurate enough
partial force. If the answer is `yes', the
multipole force is used and the walk along this branch of the tree
can be terminated; if the answer is `no', the node is ``opened'', i.e.~its
daughter nodes are considered in turn. Clearly, the multipole expansion
is in general appropriate for nodes that
are sufficiently small and distant. Most commonly one uses a fixed
angle (typically $\approx 0.5$~rad) as opening criteria.

It should be noted that the final result of the tree algorithm
will in general only represent an approximation to the true force.
However, the error can be controlled conveniently by modifying the
opening criterion for tree nodes, because a higher accuracy is
obtained by walking the tree to lower levels. Provided that sufficient
computational resources are invested, the tree force can then be
made arbitrarily close to the well-specified correct force.
Nevertheless evaluating the gravitational force via a tree
leads to an inherent asymmetry in the interaction between two
particles. It is worth mentioning that there are extensions to the
standard tree, the so-called {\sl fast multipole methods}, which avoid
these asymmetries, and therefore have better conservation of
momentum. For an N-body application of such a technique
see \citet{2000ApJ...536L..39D} and references therein. However,
these methods compute the forces for all the particles at every time step 
and can not take advantage of
using individual time steps for different particles.

\subsection{Particle-Mesh methods}

The Particle-Mesh ({\sl PM}) method treats the force as a field quantity
by computing it on a mesh. Differential operators, such as the
Laplacian, are replaced by finite difference approximations.
Potentials and forces at particle positions are obtained by
interpolation on the array of mesh-defined values. Typically, such
an algorithm is performed in three steps. First, the density on the mesh
points is computed by assigning densities to the mesh from the
particle positions. Second, the density field is transformed to
Fourier space, where the Poisson equation is solved, and the potential
is obtained using Green's method. Alternatively, the potential
can be determined by solving Poisson's equation iteratively with
relaxation methods. In a third step the forces for the individual
particles are obtained by interpolating the derivatives of the
potentials to the particle positions. Typically, the amount of mesh
cells $N$ used corresponds to the number of particles in the simulation, so
that when structures form, one can have large numbers of particles
within individual mesh cells, which immediately illustrates the
shortcoming of this method; namely its limited resolution. On the
other hand, the calculation of the Fourier transform via a Fast Fourier
Transform (FFT) is extremely fast, as it only needs of order $N
\log N$ operations, which is the advantage of this method. Note
that here $N$ denotes the number of mesh cells. 
In this approach the computational
costs do not depend on the details of the particle distribution. Also
this method can not take advantage of individual time steps, as the
forces are always calculated for all particles at every time step.

There are many schemes to assign the mass density to the mesh. The
simplest method is the ``Nearest-Grid-Point'' (NGP). Here,
each particle is assigned to the closest mesh point, and
the density at each mesh point is the total
mass assigned to the point divided by the
cell volume. However, this method is rarely used. One of its
drawbacks is that it gives forces that are discontinuous.
The ``Cloud-in-a-Cell'' (CIC) scheme is a better
approximation to the force: it distributes every particle over the
nearest 8 grid cells, and then weighs them by the overlapping volume,
which is obtained by
assuming the particle to have a cubic shape of the same volume as
the mesh cells. The CIC method gives continuous forces,
but discontinuous first derivatives of the forces. A more accurate scheme
is the ``Triangular-Shaped-Cloud'' (TSC) method. This scheme has an
assignment interpolation function that is piecewise quadratic. In
three dimensions it employs 27 mesh points
\citep[see][]{hockney88}.

In general, one can define the assignment of the density $\rho_m$
on a grid ${\vec x}_m$ with spacing $\delta$ from the distribution of particles
with masses $m_i$ and positions ${\vec x}_i$, by smoothing the particles over
$n$ times the grid spacing ($h = n\delta$). Therefore, having
defined a weighting function
\begin{equation}
   W({\vec x}_m-{\vec x}_i)= \int\hat{W}\left(\frac{{\vec x}-{\vec x_m}}{h}\right)
                           \, S({\vec x}-{\vec x_i},h) \dd{\vec x},
\end{equation}
where $\hat{W}({\vec x})$ is $1$ for $|{\vec x}|<0.5$ and $0$
otherwise, the density $\rho_m$ on the grid can be written as
\begin{equation}
   \rho_m = \frac{1}{h^3}\sum_i m_i W({\vec x}_i - {\vec x}_m).
\end{equation}
The shape function $S({\vec x},h)$ then defines the different
schemes. The aforementioned NGP, CIC and TSC schemes are equivalent to the
choice of 1, 2 or 3 for $n$ and the Dirac $\delta$ function
$\delta({\vec x})$,
$\hat{W}({\vec x}/h)$ and $1-|{\vec x}/h|$ for the shape function
$S({\vec x},h)$, respectively.

In real space, the gravitational potential $\Phi$ can be written
as the convolution of the mass density with a suitable Green's
function $g({\vec x})$:
\begin{equation}
\Phi({\vec x}) = \int g({\vec x} - {\vec x}') \rho({\vec x}') \dd
{\vec x}'.
\label{phi-green}
\end{equation}
For vacuum boundary conditions, for
example, the gravitational potential is
\begin{equation}
\Phi({\vec x}) = -{\rm G} \int
\frac{\rho({\vec x}')}{|{\vec x}-{\vec x}'|}\dd{\vec x}',
\end{equation}
with ${\rm G}$ being the gravitational constant. Therefore the Green's
function, $g({\vec x}) = -{\rm G}/|{\vec x}|$,
represents the solution
of the Poisson equation ${\vec \nabla}^2 \Phi({\vec x}) = 4\pi {\rm G} \rho({\vec
x})$, recalling that ${\vec \nabla}_x^2(|{\vec x}-{\vec x}'|)^{-1}
= 4\pi\delta({\vec x}-{\vec x}')$. By applying 
the divergence theorem to the integral form of the above equation,
it is then easy to see that, in spherical coordinates,  
\begin{equation}
\int_V {\vec \nabla}^2 \left(\frac{1}{r}\right)\dd V =
\int_S {\vec \nabla} \left(\frac{1}{r}\right)\dd S =
\int_0^{2\pi}\int_0^\pi\frac{\partial}{\partial r}
\left(\frac{1}{r}\right)r^2\mathrm{sin}(\theta)\dd\theta\dd\phi = -4\pi.
\end{equation}
Periodic boundary conditions are usually used to simulate an ``infinite
universe'', however zero padding can be applied to deal with vacuum
boundary conditions.

In the {\sl PM} method, the solution to the Poisson equation is easily
found in Fourier space, where Eq.~\ref{phi-green} becomes a simple multiplication
\begin{equation}
\hat{\Phi}({\vec k}) = \hat{g}({\vec k}) \, \hat{\rho}({\vec k}).
\end{equation}
Note that $\hat{g}({\vec k})$ has only to be computed once, at
the beginning of the simulation.

After the calculation of the potential via Fast Fourier Transform
(FFT) methods, the force
field ${\vec f}({\vec x})$ at the position of the mesh points can be obtained
by differentiating the potential, ${\vec f}({\vec x}) =
{\vec \nabla}\Phi({\vec x})$. This can be done by a finite-difference
representation of the gradient. In a second order scheme, the
derivative with respect to the $x$ coordinate at the mesh
positions $m=(i,j,k)$ can be written as
\begin{equation}
   f_{i,j,k}^{(x)} = -\frac{\Phi_{i+1,j,k}-\Phi_{i-1,j,k}}{2h}.
\label{eq:finite1}
\end{equation}
A fourth order scheme for the derivative would be written as
\begin{equation}
   f_{i,j,k}^{(x)} = -\frac{4}{3}\frac{\Phi_{i+1,j,k}-\Phi_{i-1,j,k}}{2h}
                   +\frac{1}{3}\frac{\Phi_{i+2,j,k}-\Phi_{i-2,j,k}}{4h}.
\label{eq:finite2}
\end{equation}

Finally, the forces have to be interpolated back to the particle
positions as
\begin{equation}
   {\vec f}({\vec x}_i) = \sum_m W({\vec x}_i - {\vec x}_m) {\vec f}_m,
\end{equation}
where it is recommended to use the same weighting scheme as for
the density assignment; this ensures pairwise force symmetry between
particles and momentum conservation.

The advantage of such {\sl PM} methods is the speed, because the number of
operations scales with $N+N_g{\mathrm {log}}(N_g)$, where $N$ is the number
of particles and $N_g$ the number of mesh points. However, the
disadvantage is that the dynamical range is limited by
$N_g$, which is usually limited by the available memory.
Therefore, particularly
for cosmological simulations, adaptive methods are needed to
increase the dynamical range and follow
the formation of individual objects.

\begin{figure*}
\begin{center}
\includegraphics[width=\textwidth]{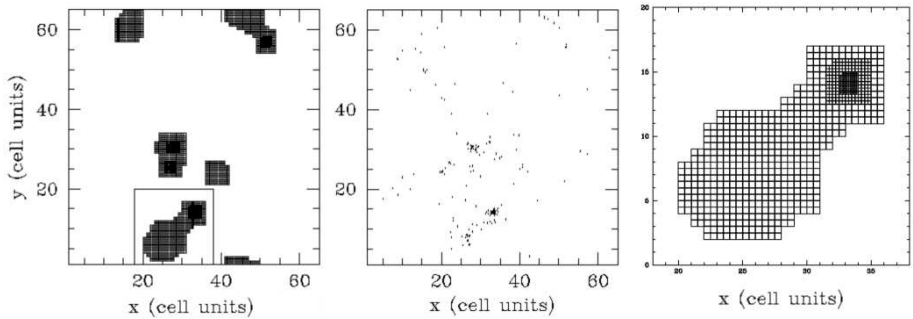}\\
\end{center}
\caption{A slice through the refinement structure (the base grid is
not shown) in a $\Lambda$CDM simulation (left panel) and the
corresponding slice through the particle distribution (middle
panel). The area enclosed by the square is enlarged in the right
panel. Taken from \protect\citet{1997ApJS..111...73K}.} \label{ART_PM}
\end{figure*}

In the Adaptive Mesh Refinement ({\sl AMR}) techniques,
the Poisson equation on the refinement
meshes can be treated as a Dirichlet boundary problem for which
the boundary values are obtained by interpolating the gravitational potential from
the parent grid. In such algorithms, the boundaries of the
refinement meshes can have an arbitrary shape; this feature narrows the
range of solvers that one can use for partial differential equation
(PDEs). The Poisson equation on these meshes can be solved using the
{\em relaxation} method \citep{hockney88,1992nrfa.book.....P},
which is relatively fast and efficient in dealing with complicated
boundaries. In this method the Poisson equation
\begin{equation}
\nabla^2 \Phi = \rho
\end{equation}
is rewritten in the form of a diffusion equation,
\begin{equation}
\frac{\partial \Phi}{\partial \tau}=\nabla^2 \Phi - \rho.
\end{equation}
The point of the method is that an initial solution guess $\Phi$
{\em relaxes} to an equilibrium solution (i.e., solution of the
Poisson equation) as $\tau \to \infty$. The finite-difference form
of Eq.~2 is:
\begin{equation}
\Phi^{n+1}_{i,j,k}=\Phi^n_{i,j,k}+\frac{\Delta \tau}{\Delta^2}
\left(\sum^6_{nb=1}\Phi^n_{nb}-6\Phi^n_{i,j,k}\right)
-\rho_{i,j,k}\Delta \tau.
\end{equation}
where the summation is performed over a cell's neighbours. Here,
$\Delta$ is the actual spatial resolution of the solution
(potential), while $\Delta \tau$ is a fictitious time step (not
related to the actual time integration of the $N$-body system).
This finite difference method is stable when $\Delta
\tau\leq\Delta^2/6$. More details can be found in Press et al.
(1992) and also \citet{1997ApJS..111...73K}. Fig.~\ref{ART_PM}, 
from \citet{1997ApJS..111...73K}, shows an example of the mesh constructed to
calculate the potential in a cosmological simulation.

\subsection{Hybrids (TreePM/P$^3$M)}

\begin{figure}
\begin{center}
\includegraphics[width=0.41\textwidth]{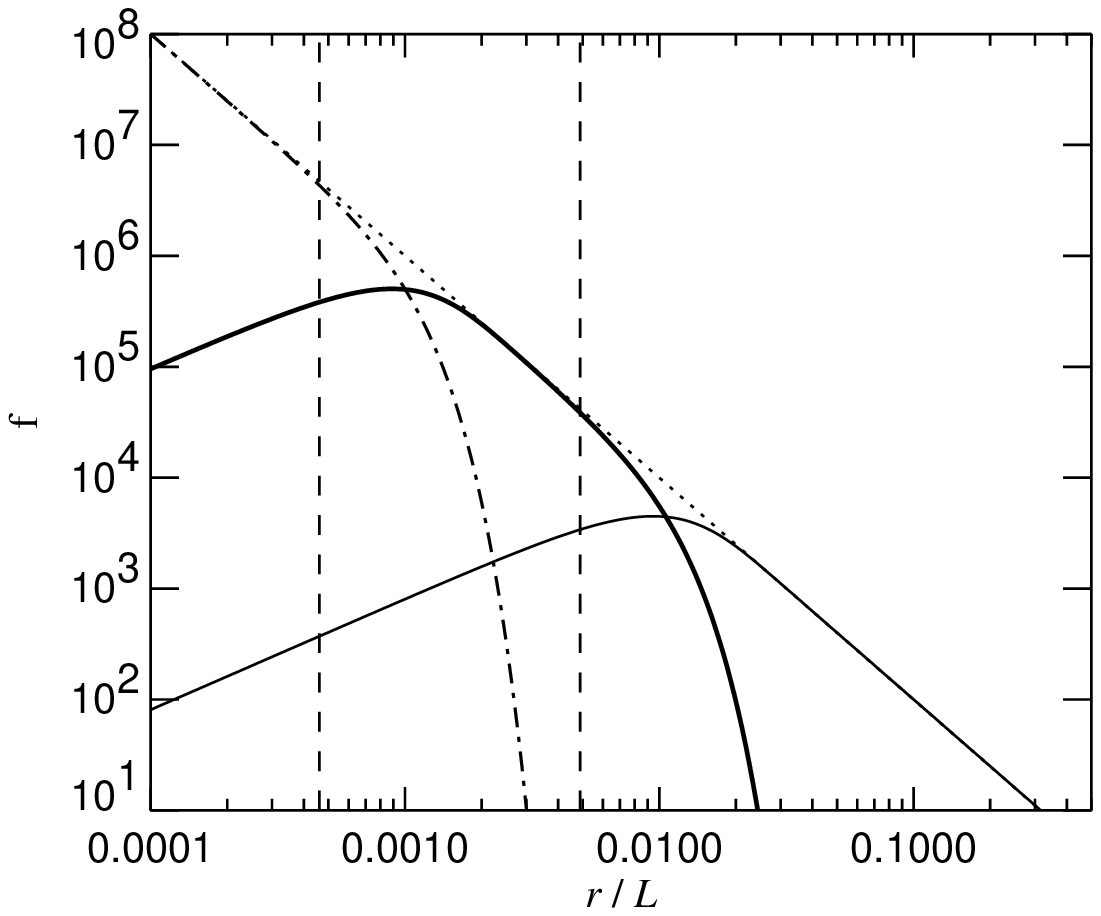}
\includegraphics[width=0.54\textwidth]{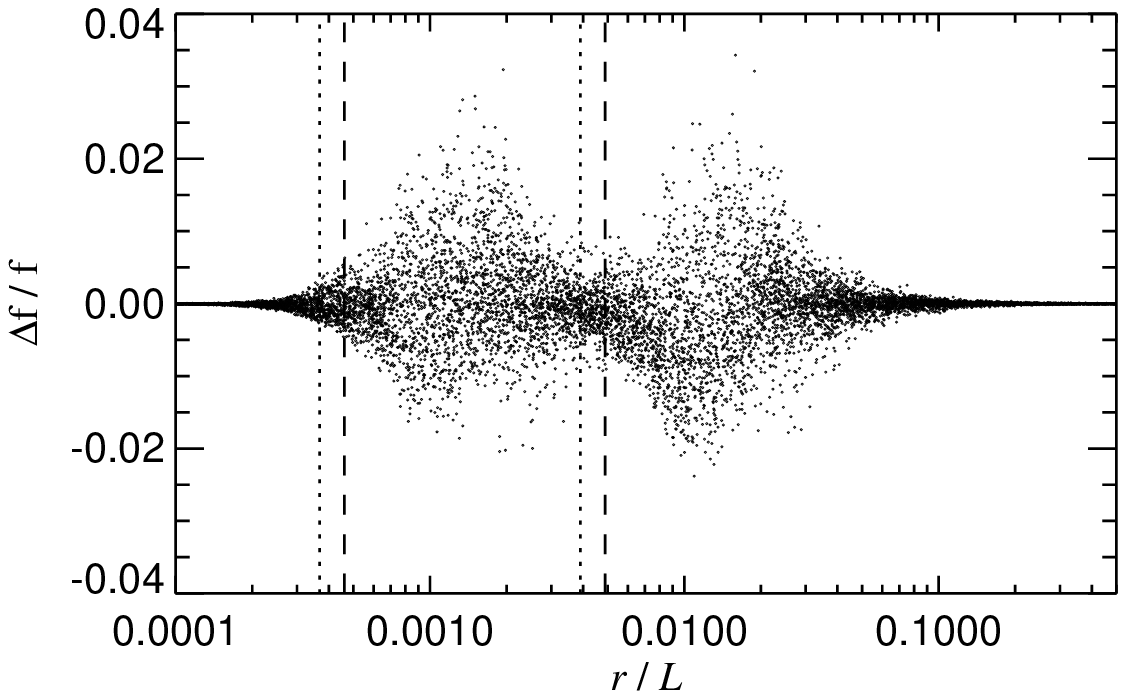}
\end{center}
\caption{Force decomposition and force error of the {\sl TreePM} scheme
in the case when two stacked meshes are used. The left panel
illustrates the strength of the short-range (dot-dashed),
intermediate-range (thick solid), and long-range (solid) force as
a function of distance in a periodic box. The spatial scales of
the two splits are marked with vertical dashed lines. The right
panel shows the error distribution of the PM force. The outer
matching region exhibits a very similar error characteristic as
the inner match of tree- and PM-force. In both cases, for
separations of order the fine or coarse mesh scale (dotted lines),
respectively, force errors of up to $1-2$~\% arise, but the
r.m.s. force error stays well below 1~\%, and the mean force
tracks the correct result accurately. Taken from
\citet{springel2005}.} \label{force_match}
\end{figure}

Hybrid methods can be constructed as a synthesis of the
particle-mesh method and the tree algorithm. In {\sl TreePM} methods
\citep{1995ApJS...98..355X,2000ApJS..128..561B,2002JApA...23..185B,2003NewA....8..665B}
the potential is explicitly split in Fourier space into a
long-range and a short-range part according to $\Phi_{{\vec k}} =
\Phi^{{\rm long}}_{{\vec k}} + \Phi^{{\rm short}}_{{\vec k}}$, where
\begin{equation}
\Phi^{{\rm long}}_{{\vec k}} = \Phi_{{\vec k}} \exp(-{\vec k}^2
r_{\rm s}^2),
\end{equation}
with $r_{\rm s}$ describing the spatial scale of the force-split. The
long range potential can be computed very efficiently with
mesh-based Fourier methods.

The short-range part of the potential can be solved in real space
by noting that for $r_{\rm s} \ll L$ the short-range part of the
real-space solution of the Poisson equation is given by

\begin{equation}
\Phi^{{\rm short}}({\vec x}) = - G \sum_i \frac{m_i}{{\vec r}_i} {\rm
erfc}\left(\frac{{\vec r}_i} {2 r_{\rm s}}\right).
\end{equation}

Here ${\vec r}_i$ is the distance of any particle $i$ to the point
${\vec x}$. Thus the short-range force can be computed by the tree
algorithm, except that the force law is modified by a long-range
cut-off factor.

Such hybrid methods can result in a very substantial improvement
of the performance compared to ordinary tree methods.
In addition one typically gains accuracy in the long-range force,
which is now basically exact, and not an
approximation as in the tree method. Furthermore, if $r_{\rm s}$ is
chosen to be slightly larger than the mesh scale, force anisotropies,
that exist in plain {\sl PM} methods, can be suppressed to essentially
arbitrarily low levels. A {\sl TreePM} approach also maintains all
the most important advantages of the tree algorithm, namely its
insensitivity to clustering, its essentially unlimited dynamical
range, and its precise control of the softening scale of the
gravitational force.

Fig.~\ref{force_match} shows how the force matching works in
the {\sl GADGET-2} code \citep{springel2005}, where such a hybrid method
is further extended to two subsequent stacked FFTs combined with the tree
algorithm. This extension enables one to increase the dynamic range,
which, in turn, improves the computational speed in high
resolution simulations of the evolution of galaxy clusters within a
large cosmological volume.

Although used much earlier (because much easier to implement), the
{\sl P$^3$M} method can be seen as a special case of the {\sl TreePM}, where the
tree is replaced by the direct sum. Note that also in the tree
algorithm the nearest forces are calculated by a direct sum, thus the
{\sl P$^3$M} approach formally corresponds to extending the direct sum of
the tree method to the scale where the PM force computation takes
over.  \citet{1991ApJ...368L..23C} presented an improved version
of the {\sl P$^3$M} method, by allowing spatially adaptive mesh
refinements in regions with high particle density (Adaptive {\sl P$^3$M}
or {\sl AP$^3$M}). The improvement in the performance made it very
attractive and several cosmological simulations were performed with
this technique, 
including the Hubble Volume simulations \citep{2002ApJ...573....7E}.

\subsection{Time-stepping and integration}

\begin{figure}
\begin{center}
\includegraphics[width=0.45\textwidth]{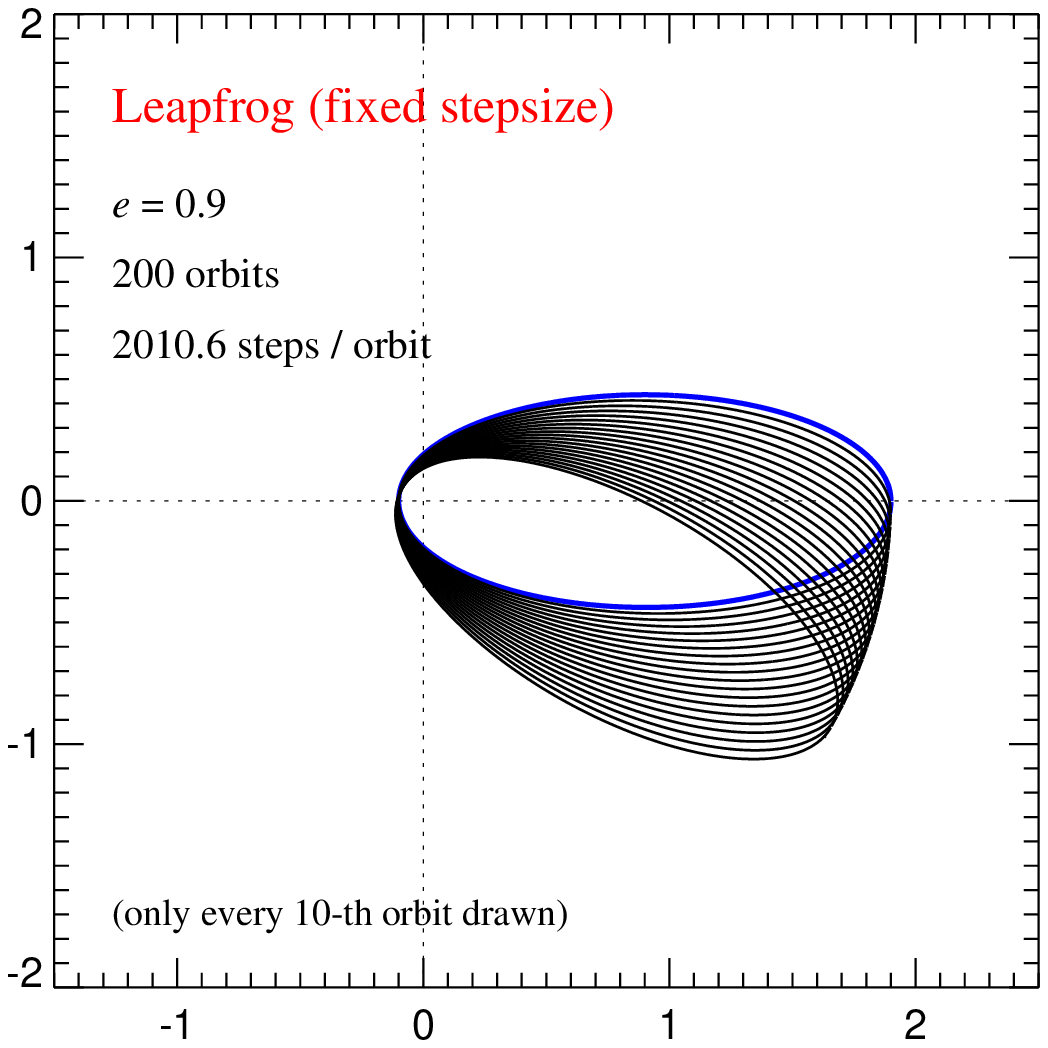}
\includegraphics[width=0.45\textwidth]{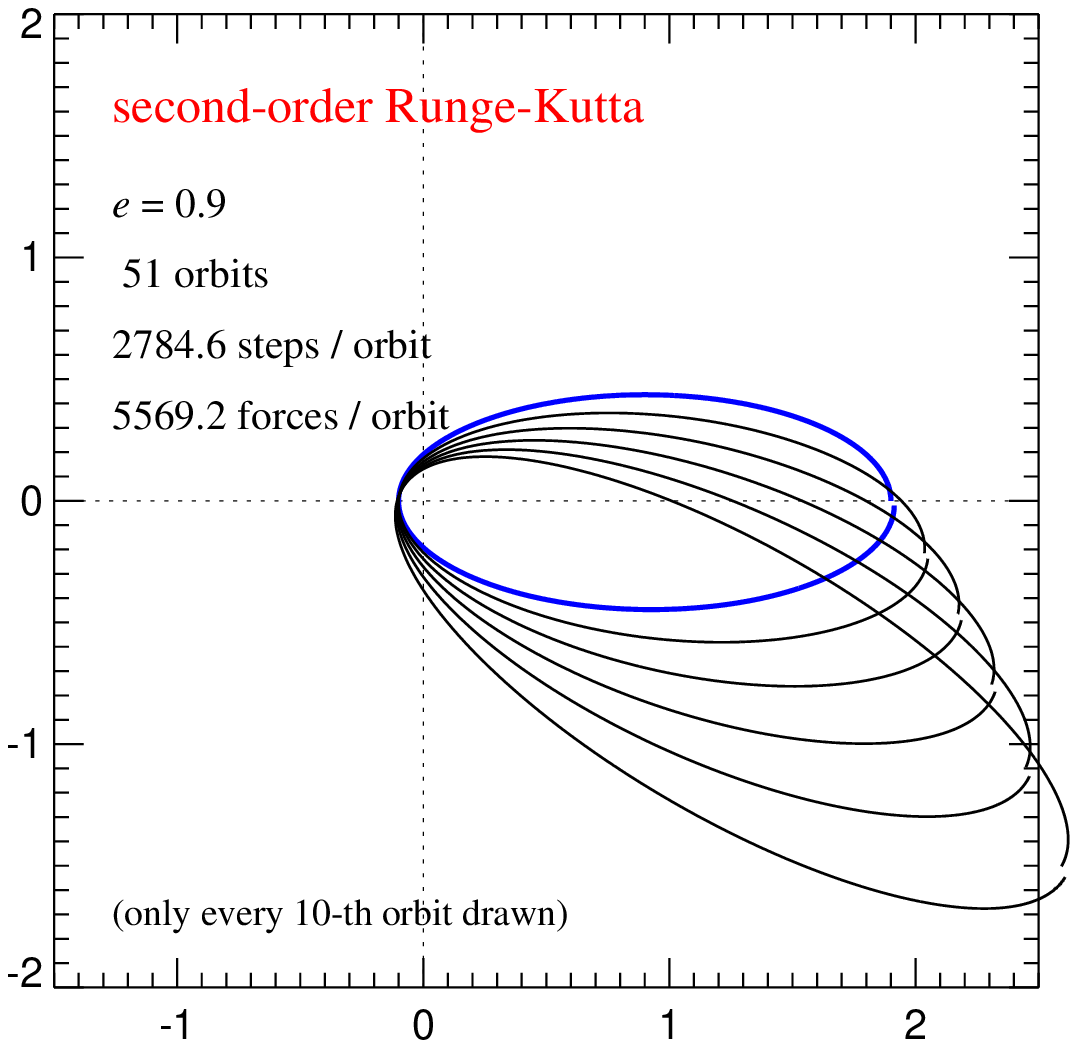}
\includegraphics[width=0.45\textwidth]{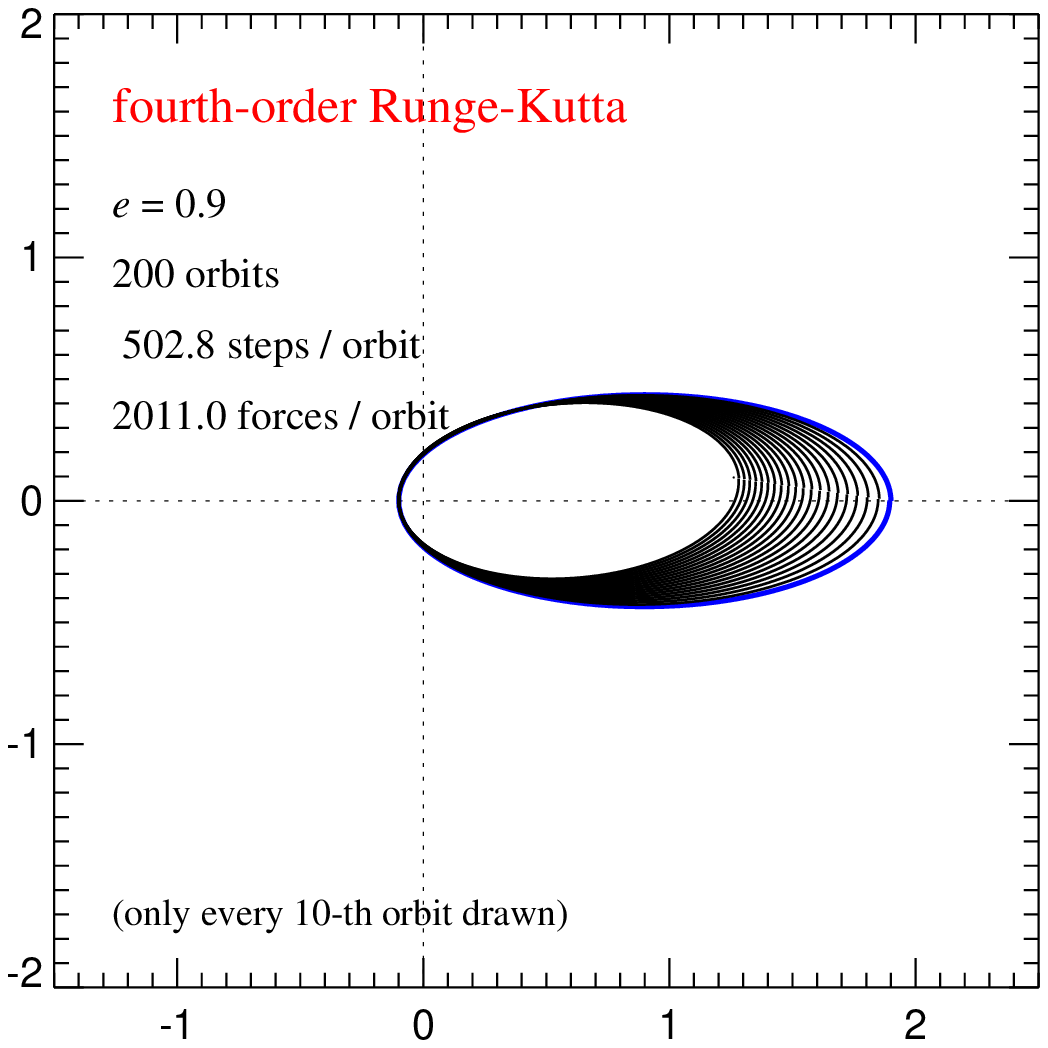}\\
\includegraphics[width=0.45\textwidth]{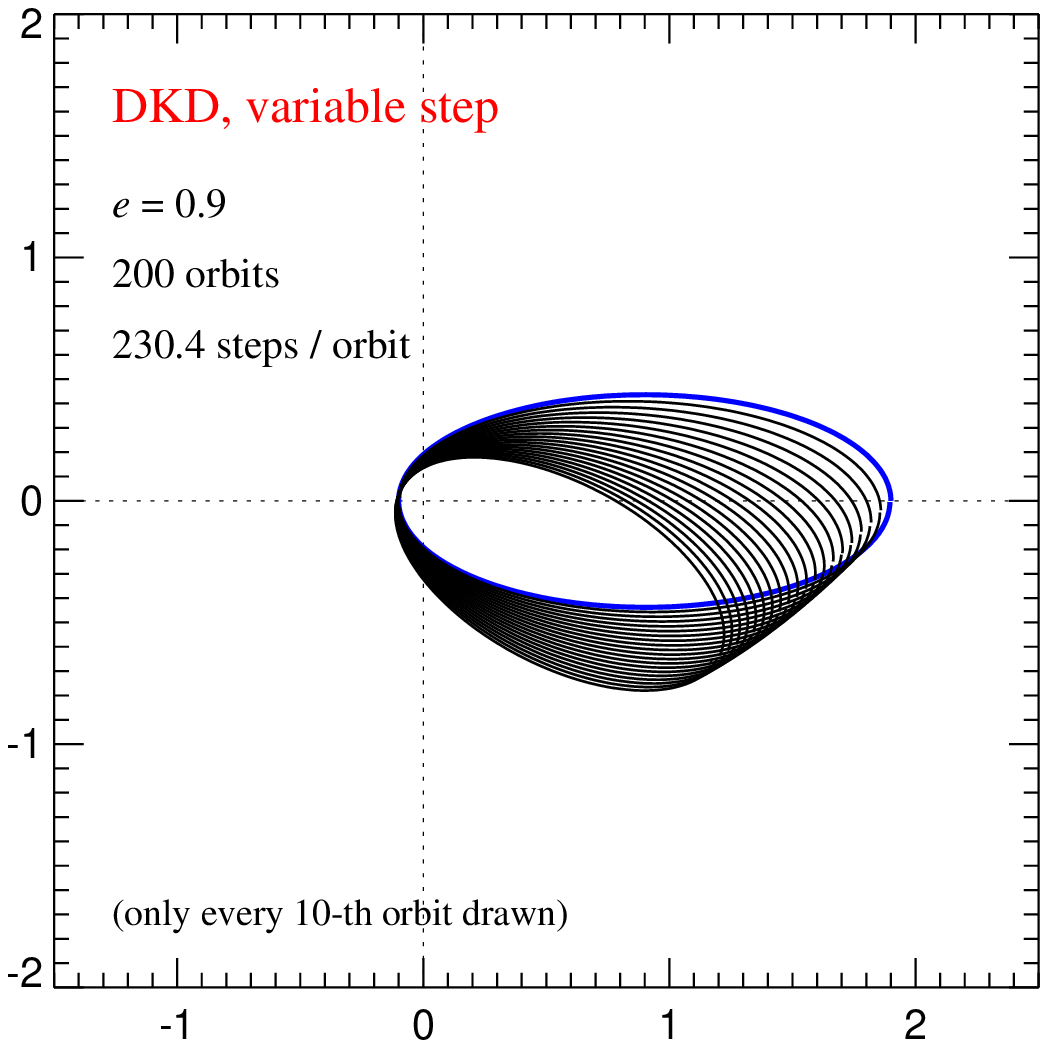}
\includegraphics[width=0.45\textwidth]{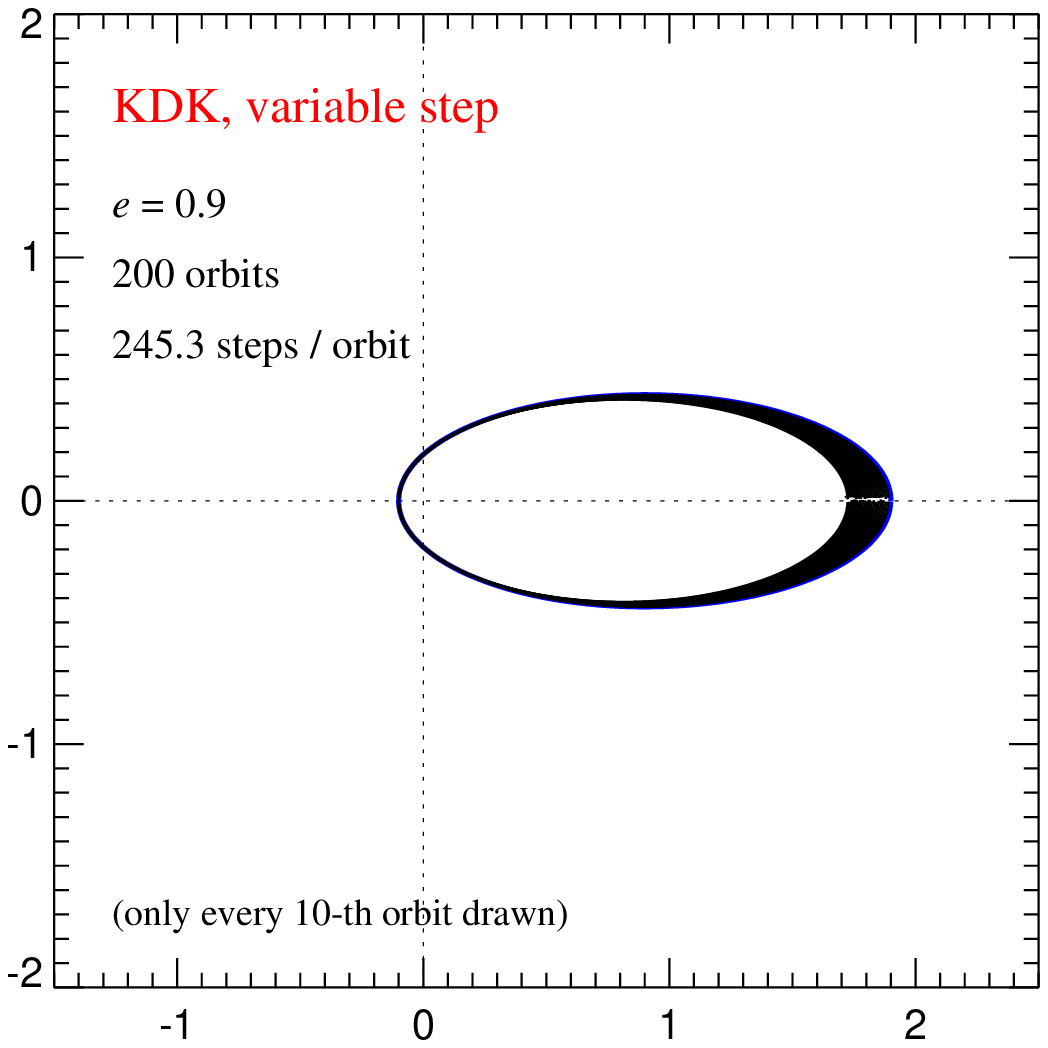}
\end{center}
\caption{The upper two rows shows a Kepler problem of high eccentricity
evolved with different simple time integration schemes, using an
equal time step in all cases. Even though the leap-frog (upper left
panel) and the second order Runge-Kutta (upper right panel) produce
comparable errors in a single step, the long term stability of the
integration is very different. Even a computationally much more
expensive fourth order Runge-Kutta scheme (middle row), with a
smaller error per step, performs dramatically worse than the
leap-frog in this problem. The lower row shows the same using leap-frog
schemes with a variable time step from step to step, based on the
$\Delta t \propto 1/\sqrt{|{\vec a}|}$ criterion commonly employed
in cosmological simulations. As a result of the variable time steps, 
the integration is no longer manifestly time reversible,
and long term secular errors develop. Interestingly, the error in
the KDK (Kick-Drift-Kick) variant grows four times more slowly than in 
the DKD (Drift-Kick-Drift) variant,
despite being of equal computational cost. Taken from
\protect\citet{springel2005}. } \label{fig_timestep}
\end{figure}

The accuracy obtained when evolving the system depends on the size
of the time step and on the integrator scheme used. Finding the
optimum size of time step is not trivial. A very simple criterion
often used is
\begin{equation}
   \Delta t = \alpha \sqrt{\epsilon / |{\vec a}|}
\end{equation}
where $|a|$ is the acceleration obtained at the previous time step, 
$\epsilon$ is a length scale, which can typically be
associated with the gravitational softening, and $\alpha$ is a
tolerance parameter. More details about different time step criteria
can be found for example in \citet{2003MNRAS.338...14P} and references
therein. For the integration of the variables (positions and
velocities) of the system, we only need to integrate first order
equations of the form $\dot{{\vec y}}=f({\vec y})$, e.g. ordinary
differential equations (ODEs) with appropriate initial conditions.
Note that we can first solve this ODE for the velocity ${\vec v}$
and then treat $\dot{{\vec x}} = {\vec v}$ as an independent ODE, at
basically no extra cost.

One can distinguish implicit and explicit methods for propagating
the system from step $n$ to step $n+1$. Implicit methods
usually have better properties, however they need to solve
the system iteratively, which usually requires inverting
a matrix which is only sparsely sampled, and has the dimension
of the total number of the data points, namely grid or particle points.
Therefore, N-body simulations mostly adopt explicit methods.

The simplest (but never used) method to perform the integration of
an ODE is called Euler's method; here the integration is
just done by multiplying the derivatives  with the length of the
time step. The explicit form of such a method can be written as
\begin{equation}
   {\vec y}_{n+1} = {\vec y}_n + {\vec f}({\vec y}_n)\Delta t,
\end{equation}
whereas the implicit version is written as
\begin{equation}
   {\vec y}_{n+1} = {\vec y}_n + {\vec f}({\vec y}_{n+1})\Delta t.
\end{equation}
Note that in the latter equation ${\vec y}_{n+1}$ appears on the left
and right side, which makes it clear why it is called
implicit. Obviously the drawback of the explicit method is that it
assumes that the derivatives (e.g. the forces) do not change during
the time step.

An improvement to this method can be obtained by using the
mean derivative during the time step, which can be written with the
{\sl implicit mid-point rule} as
\begin{equation}
   {\vec y}_{n+1} = {\vec y}_n + {\vec f}[0.5({\vec y}_{n}+{\vec y}_{n+1})]\Delta t.
\end{equation}
An explicit rule using the forces at the next time step is the
so-called {\sl predictor-corrector method}, where one first predicts
the variables for the next time step
\begin{equation}
   {\vec y}_{n+1}^0 = {\vec y}_n + {\vec f}({\vec y}_n)\Delta t
\end{equation}
and then uses the forces calculated there to correct this
prediction (the so-called corrector step) as
\begin{equation}
   {\vec y}_{n+1} = {\vec y}_n + 0.5[{\vec f}({\vec y}_n)+{\vec f}({\vec y}_{n+1}^0)]\Delta
   t.
\end{equation}
This method is accurate to second order.

In fact, all these methods are special cases of the so-called {\sl
Runge-Kutta method} (RK), which achieves the accuracy of a Taylor
series approach without requiring the calculation of  higher order
derivatives. The price one has to pay is that the derivatives
(e.g. forces) have to be calculated at several points, effectively
splitting the interval $\Delta t$ into special subsets. For example,
a second order RK scheme can be constructed by
\begin{equation}
{\vec k}_1 = {\vec f}({\vec y}_n)
\end{equation}
\begin{equation}
{\vec k}_2 = {\vec f}({\vec y}_n+{\vec k}_1\Delta t)
\end{equation}
\begin{equation}
{\vec y}_{n+1} = {\vec y}_n + 0.5({\vec k}_1+{\vec k}_2)\Delta t.
\end{equation}
In a fourth order RK scheme, the time interval $\Delta t$ also has
to be subsampled to calculate the mid-points, e.g.
\begin{equation}
{\vec k}_1 = {\vec f}({\vec y}_n,t_n)
\end{equation}
\begin{equation}
{\vec k}_2 = {\vec f}({\vec y}_n+{\vec k}_1\Delta t/2,t_n+\Delta t/2)
\end{equation}
\begin{equation}
{\vec k}_3 = {\vec f}({\vec y}_n+{\vec k}_2\Delta t/2,t_n+\Delta t/2)
\end{equation}
\begin{equation}
{\vec k}_4 = {\vec f}({\vec y}_n+{\vec k}_3\Delta t/2,t_n+\Delta t)
\end{equation}
\begin{equation}
{\vec y}_{n+1} = {\vec y}_n + \left(\frac{{\vec k}_1}{6} +
\frac{{\vec k}_2}{3} + \frac{{\vec k}_3}{3} + \frac{{\vec k}_4}{6}
\right)\Delta t.
\end{equation}
More details on how to construct the coefficient for an $n$-th order
RK scheme are given in e.g. \citet{NumMeth550530}.

Another possibility is to use the so-called {\sl leap-frog
method}, where the derivatives (e.g. forces) and
the positions are shifted in time by
half a time step. This feature can be
used to integrate directly the second order ODE of the form
$\ddot{{\vec x}} = {\vec f}({\vec x})$. Depending on whether one starts with a
{\sl drift (D}) of the system by half a time step or one uses
the forces at the actual time to propagate the system ({\sl kick, K}), 
one obtains a KDK version
\begin{equation}
{\vec v}_{n+1/2} = {\vec v}_n + {\vec f}({\vec x}_n)\Delta t/2
\end{equation}
\begin{equation}
{\vec x}_{n+1} = {\vec x}_n + {\vec v}_{n+1/2}\Delta t
\end{equation}
\begin{equation}
{\vec v}_{n+1} = {\vec v}_{n+1/2} + {\vec f}({\vec x}_{n+1})\Delta t/2
\end{equation}
or a DKD version of the method
\begin{equation}
{\vec x}_{n+1/2} = {\vec x}_n + {\vec v}_{n}\Delta t/2
\end{equation}
\begin{equation}
{\vec v}_{n+1} = {\vec v}_n + {\vec f}({\vec x}_{n+1/2})\Delta t
\end{equation}
\begin{equation}
{\vec x}_{n+1} = {\vec x}_{n+1/2} + {\vec v}_{n+1}\Delta t/2.
\end{equation}
This method is accurate to second order, and, as will be shown in
the next paragraph, also has other advantages. For more details see
\citet{springel2005}.

It is also clear that, depending on the application, a lower order
scheme applied with more, and thus smaller, time steps can be more
efficient than a higher order scheme, which enables the use of
larger time steps. In the upper rows of Fig.~\ref{fig_timestep}, we show
the numerical integration of a Kepler problem
(i.e. two point-like masses with large mass difference which orbit
around each other like a planet-sun system) of high eccentricity $e=0.9$, using
second-order accurate leap-frog and Runga-Kutta schemes with fixed
time step. There is no long-term drift in the orbital energy for
the leap-frog result (left panel); only a small residual precession
of the elliptical orbit is observed. On the other hand, the second-order
Runge-Kutta integrator, which has formally the same error
per step, fails catastrophically for an equally large time step
(middle panel).  After only 50 orbits, the binding energy has
increased by $\sim 30$~\%. If we instead employ a fourth-order
Runge-Kutta scheme using the same time step (right panel), the
integration is only marginally more stable, now giving a decline
of the binding energy of $\sim 40$~\% over 200 orbits. Note however
that such a higher order integration scheme requires several force
evaluations per time step, making it computationally much more
expensive for a single step than the leap-frog, which requires only
one force evaluation per step. The underlying mathematical reason
for the remarkable stability of the leap-frog integrator lies in
its symplectic properties. For a more detailed
discussion, see \citet{springel2005}.

In cosmological simulations, we are confronted with a large
dynamic range in timescales. In high-density regions, like at the
centres of galaxies, the required time steps are orders of magnitude smaller
than in the low-density regions of the intergalactic medium,
where a large fraction of the mass resides. Hence, evolving all the particles
with the smallest required time step implies a substantial
waste of computational resources. An integration scheme with
individual time steps tries to cope with this situation more
efficiently. The principal idea is to compute forces only for a
certain group of particles in a given kick operation (K), with the
other particles being evolved on larger time steps being usually just
drifted (D) and `kicked' more rarely.

The KDK scheme is hence clearly superior once one allows for
individual time steps, as shown in the lower row of Fig.~\ref{fig_timestep}.
It is also possible to try to recover the time
reversibility more precisely. \cite{1995ApJ...443L..93H} discuss
an implicit time step criterion that depends both on the beginning
and on the end of the time step, and, similarly,
\cite{1997astro.ph.10043Q} discuss a binary hierarchy of trial
steps that serves a similar purpose. However, these schemes are
computationally impractical for large collisionless systems.
Fortunately, however, in this case, the danger of building up large errors by
systematic accumulation over many periodic orbits is much smaller,
because the gravitational potential is highly time-dependent and
the particles tend to make comparatively few orbits over a Hubble
time.

\subsection{Initial conditions}

\begin{figure}
\begin{center}
\includegraphics[width=0.45\textwidth]{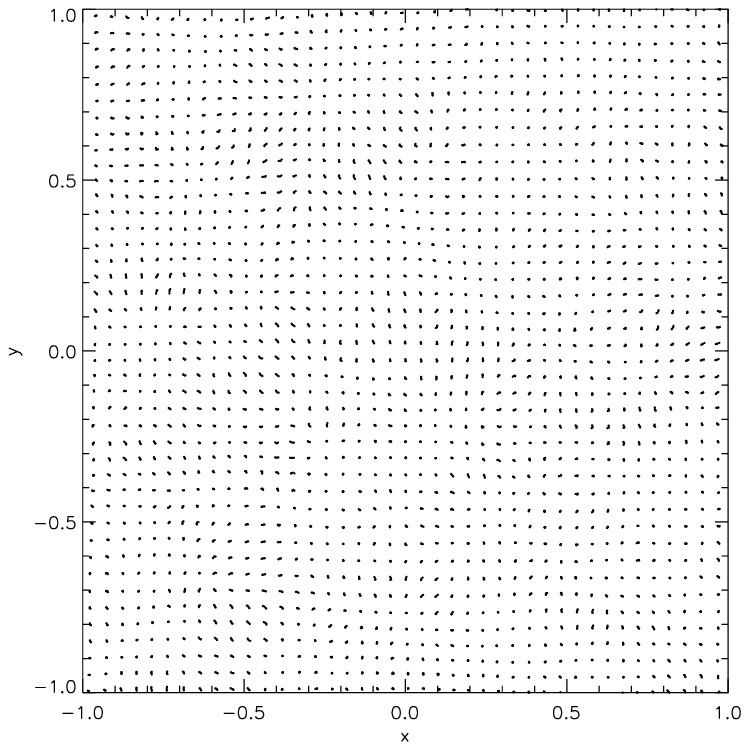}
\includegraphics[width=0.45\textwidth]{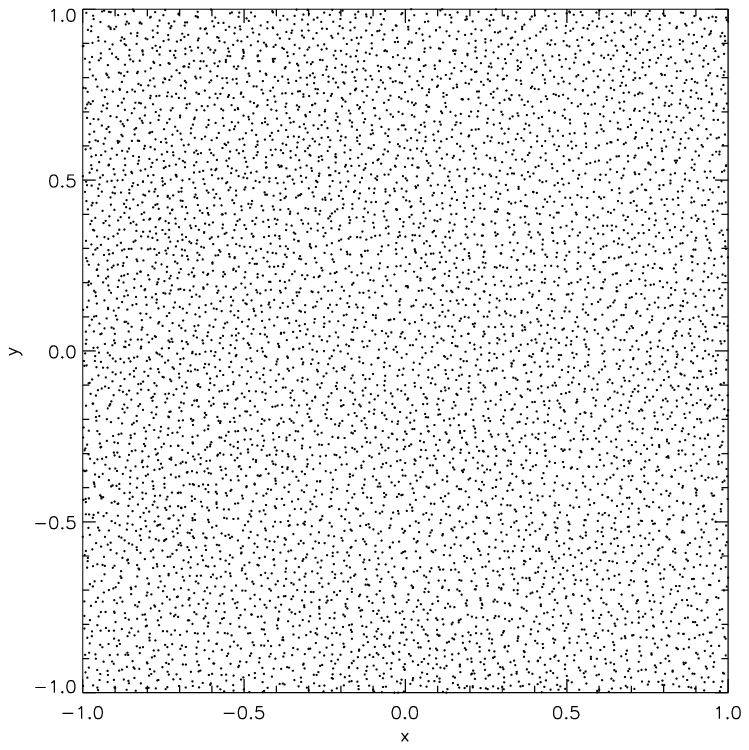}
\end{center}
\caption{Shown is a slice to the particle distribution with the
imposed displacement, taken from the same cosmological initial
conditions, once based on an originally regular grid (left panel) and
once based on an originally glass like particle distribution (right panel).
} \label{fig_ics}
\end{figure}

Having robust and well justified initial conditions is one of the key
points of any numerical effort. For cosmological purposes,
observations 
of the large--scale distribution of galaxies and of the CMB  
agree to good precision with the
theoretical expectation that the growth of structures starts from
a Gaussian random field of initial density fluctuations; this
field is thus completely described by the power spectrum $P(|{\vec k}|)$
whose shape is theoretically well motivated and depends on the
cosmological parameters and on the nature of Dark Matter.

To generate the initial conditions, one has to generate a
set of complex numbers with a randomly distributed phase $\phi$
and with amplitude normally distributed with a variance given by
the desired spectrum \citep[e.g.][]{1986ApJ...304...15B}. This can
be obtained by drawing two random numbers $\phi$ in $]0,1]$ and
$A$ in $]0,1]$ for every point in $k$-space
\begin{equation}
   \hat{\delta}_{{\vec k}} =
   \sqrt{-2P(|{\vec k}|)\mathrm{ln}(A)}{\rm e}^{i2\pi\phi}.
\end{equation}
To obtain the perturbation field generated from this distribution,
one needs to generate the potential $\Phi({\vec q})$ on a grid
${\vec q}$ in real space via a Fourier transform, e.g.
\begin{equation}
   \Phi({\vec q}) = \sum_k
   \frac{\hat{\delta}_{{\vec k}}}{{\vec k}^2}{\rm e}^{i{\vec k}{\vec q}}.
\end{equation}
The subsequent application of the Zel'dovich approximation
\citep{1970A&A.....5...84Z} enables one to find the initial
positions
\begin{equation}
{\vec x} = {\vec q} - D^+(z)\Phi({\vec q})
\end{equation}
and velocities
\begin{equation}
{\vec v} = \dot{D}^+(z) {\vec \nabla}\Phi({\vec q})
\end{equation}
of the particles, where $D^+(z)$ and $\dot{D}^+(z)$ indicate the
cosmological linear growth factor and its derivative at the initial
redshift $z$. A more detailed description can be found in e.g.
\citet{1985ApJS...57..241E}.

There are two further complications which should be mentioned. The
first is that one can try to reduce the discreteness effect
that is induced on the density power spectrum by the regularity of the
underlying grid of the particle positions ${\vec q}$ that one has at
the start. This can be done by constructing an amorphous, fully relaxed
particle distribution to be used, instead of a regular grid. Such a
particle distribution can be constructed by applying negative
gravity to a system and evolving it for a long time, including a
damping of the velocities, until it reaches a relaxed state, as
suggested by \citet{1996clss.conf..349W}. Fig.~(\ref{fig_ics}) gives a visual
impression on the resulting particle distributions.

A second complication is that, even for studying individual
objects like galaxy clusters, large-scale tidal forces can be
important. A common approach used to deal with this problem is the so-called ``zoom''
technique: a high resolution region is self-consistently
embedded in a larger scale cosmological volume at low resolution
\citep[see e.g.][]{tormen97}. This approach usually allows an increase of the dynamical
range of one to two orders of magnitude while keeping the full
cosmological context. For galaxy simulations it is even possible
to apply this technique on several levels of refinements to
further improve the dynamical range of the simulation
\citep[e.g.][]{2003MNRAS.345.1313S}. A frequently used, publicly
available package to create initial conditions is the {\sl COSMICS}
package by \citet{1995astro.ph..6070B}.

\begin{figure*}
\begin{center}
\includegraphics[width=0.6\textwidth]{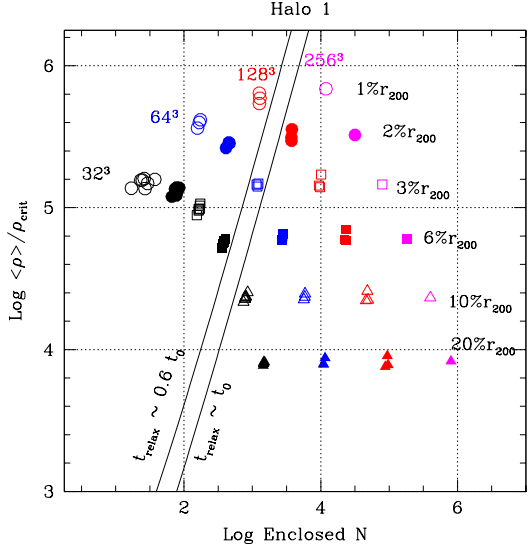}
\end{center}
\caption{Mean inner density contrast as a function of the enclosed
number of particles in 4 series of simulations varying the number
of particles in the high-resolution box, from $32^3$ to $256^3$.
Each symbol corresponds to a fixed fraction of the virial
radius, as shown by the labels on the right. The number of
particles needed to obtain robust results increases with density
contrast, roughly as prescribed by the requirement that the
collisional relaxation timescale should remain longer than the age
of the Universe. According to this, robust numerical estimates of
the mass profile of a halo are only possible to the right of the
curve labelled $t_{\rm relax}\sim 0.6 t_0$. Taken from
\protect\citet{2003MNRAS.338...14P}.} \label{figs:rrho_nenc}
\end{figure*}

\subsection{Resolution}

There has been a long standing discussion in the literature to understand
what is the optimal setup for cosmological simulations, and how
many particles are needed to resolve certain regions of interest.
Note that the number of particles needed for convergence also depends
on what quantity one is interested in. For example, mass
functions, which count identified haloes, usually give
converging results at very small particle numbers per halo
($\approx 30-50$), whereas structural properties, like a central
density or the virial radius, converge only at significantly
higher particle numbers ($\approx 1000$). As we will see in a
later chapter, if one wants to infer hydrodynamical properties
like baryon fraction or X-ray luminosity, values converge only for
haloes represented by even more particles ($\approx 10\,000$).

Recently, \citet{2003MNRAS.338...14P} performed a comprehensive
series of convergence tests designed to study the effect of
numerical parameters on the structure of simulated CDM haloes.
These tests explore the influence of the gravitational softening,
the time stepping algorithm, the starting redshift, the accuracy
of force computations, and the number of particles in the
spherically-averaged mass profile of a galaxy-sized halo in the
CDM cosmogony with a non-null cosmological constant ($\Lambda$CDM).
\citet{2003MNRAS.338...14P}, and the references therein,
suggest empirical rules that optimise the choice of these parameters.
When these choices are dictated by computational limitations, \citet{2003MNRAS.338...14P}
offer simple prescriptions to assess the effective convergence of
the mass profile of a simulated halo. One of their
main results is summarised in Fig.~\ref{figs:rrho_nenc}, which
shows the convergence of a series of simulations with different
mass resolution on different parts of the density profile of a
collapsed object. This figure clearly demonstrates that the number of
particles within a certain radius needed to obtain converging
results depends on the enclosed density.

\begin{figure*}
\begin{center}
\includegraphics[width=\textwidth]{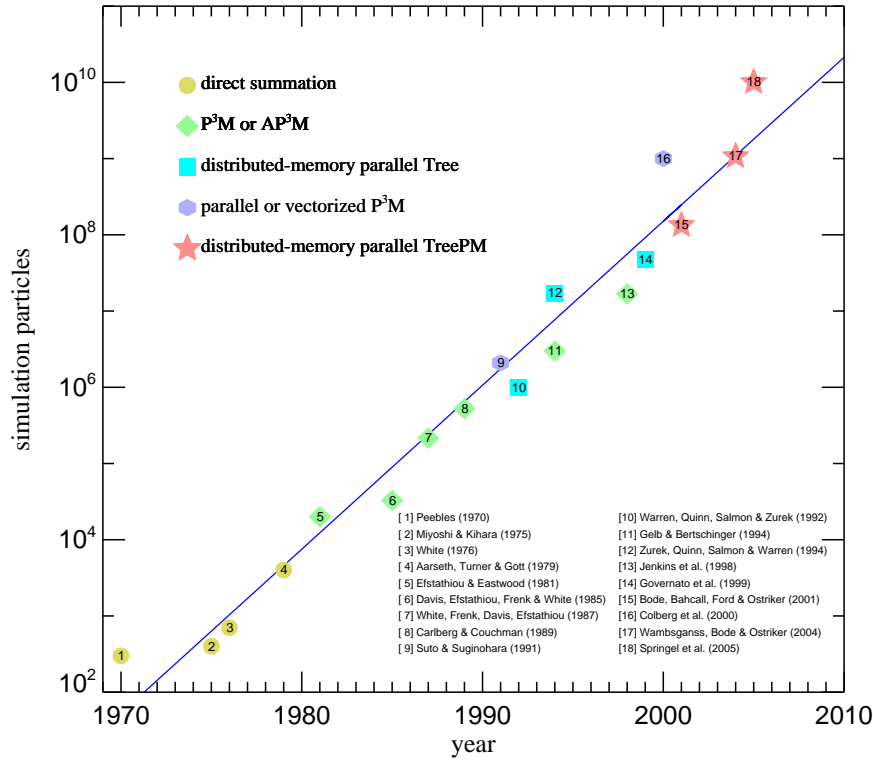}
\end{center}
\caption{Moore's empirical law shows that the computing power typically
doubles every 18 months. This figure shows the size of N-body simulations
as a function of their running date. Clearly,
specially recently, the improvement in the algorithms allowed the simulation
to grow faster than the improvement of the underlying CPU power.
Kindly provided by Volker Springel.} \label{figs:moor_nbody}
\end{figure*}

In general, both the size and the dynamical range or resolution
of the simulations have been increasing very rapidly over the last decades.
Fig.~\ref{figs:moor_nbody} shows a historical compilation of
large N-body simulations: their size growth,
thanks to improvements in the algorithms, is faster
than the underlying growth of the available CPU power.

\subsection{Code comparison for pure gravity}

\begin{figure*}
\includegraphics[width=0.49\textwidth]{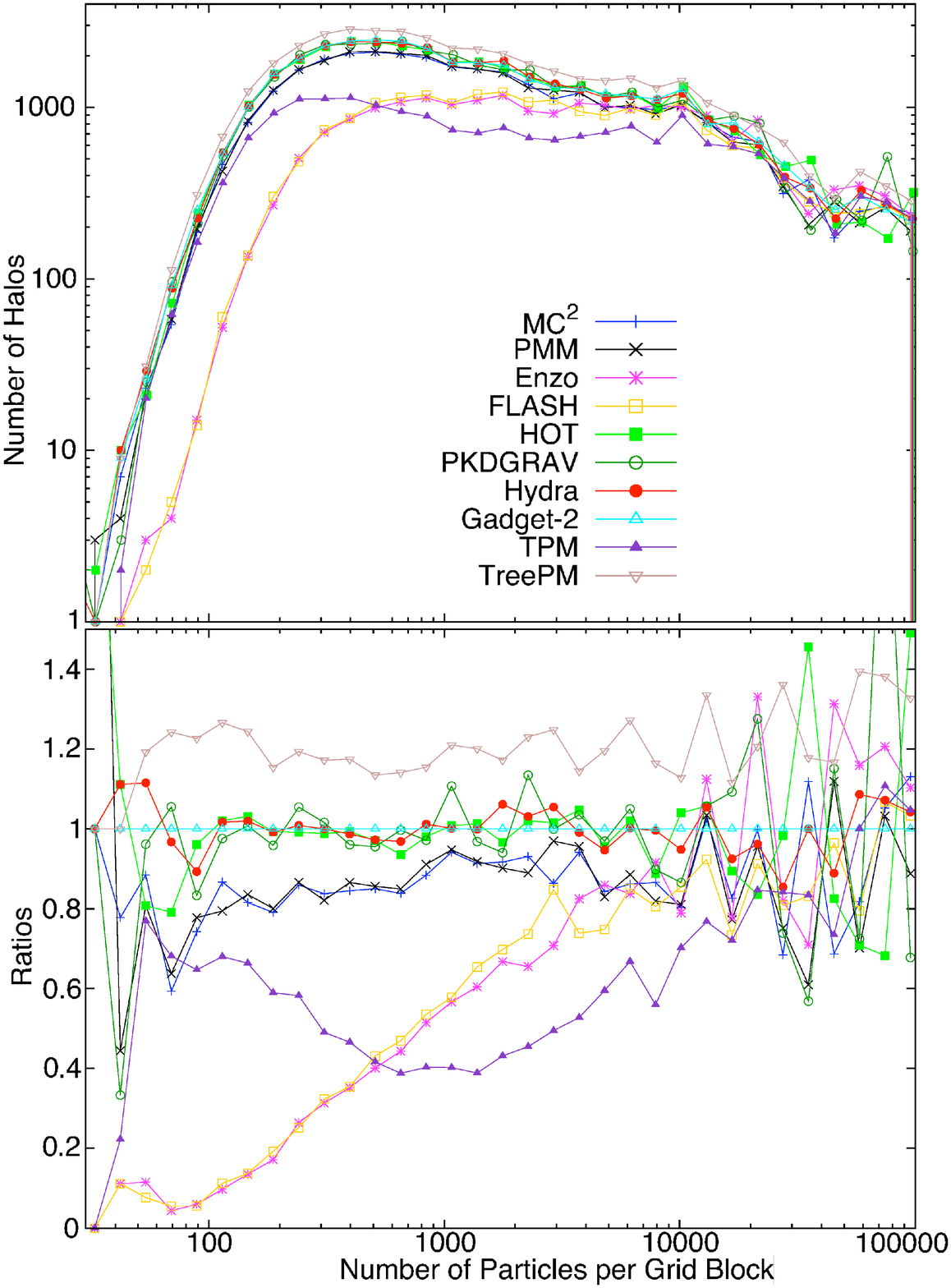}
\hspace{0.1cm}
\includegraphics[width=0.49\textwidth]{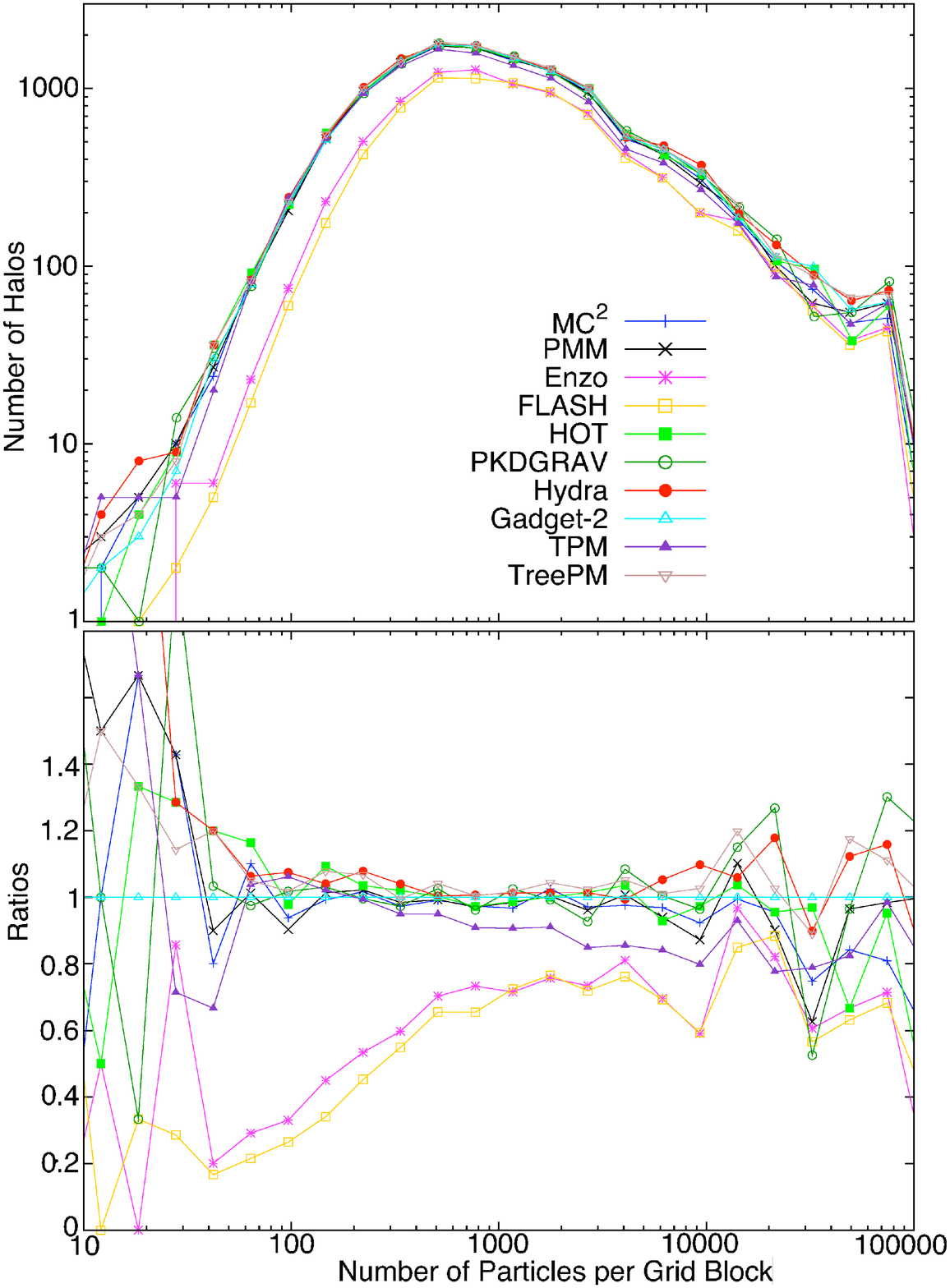}
\caption{A recent comparison of the predicted number of halos as a function of
density for ten different cosmological codes. Left panel: halos with $10 - 40$
particles, right panel: halos with $41 - 2500$ particles. The lower panels show
the residuals with respect to {\sl GADGET-2}. Both panels show the deficit of
small halos in {\sl ENZO} and {\sl FLASH} over most of the density region -- 
only at very high densities do the results catch up. The behaviour of the {\sl
TPM} simulation is interesting: not only does this simulation have a deficit of
small halos but the deficit is very significant in medium density regions, in
fact falling below the two Adaptive Mesh Refinement codes. The slight excess of
small halos shown in the {\sl TreePM} run vanishes completely if the halo cut is
raised to 20 particles per halo and the {\sl TreePM} results are in that case in
excellent agreement with {\sl GADGET-2}. Adapted from
\citet{2007arXiv0706.1270H}.}
\label{figs:halo_comp}
\end{figure*}

In the last thirty years cosmology has turned from a science of
order-of-magnitude estimates to a science with accuracies of 10~\% or
less in its measurements and theoretical predictions. Crucial
observations along the way were the measurement of the cosmic microwave
background radiation, and large galaxy surveys. In the future such
observations will yield even higher accuracy (1~\%) over a much
wider dynamical range. Such measurements will provide insight into
several topics, e.g. the nature of dark energy (expressed by the
equation of state $w=p/\rho$ with $p$ being the pressure and $\rho$ the
density). In order to make optimal use of the observations, theoretical
calculations of at least the same level of accuracy are required. As
physics in the highly non-linear regime, combined with complicated gas physics and
astrophysical feedback processes are involved, this represents a real
challenge.

Different numerical methods have therefore to be checked and
compared continuously. The most recent comparison of ten commonly-used
codes from the literature has been performed in an extensive
comparison program. The ten codes used for the comparison
performed by \citet{2007arXiv0706.1270H} cover a variety of
methods and application arenas. The simulation methods employed
include parallel particle-in-cell ({\sl PIC}) techniques (the {\sl PM} codes
{\sl MC$^2$} and {\sl PMM}, the Particle-Mesh / Adaptive Mesh Refinement ({\sl AMR})
codes {\sl ENZO} and {\sl FLASH}), a hybrid of {\sl PIC}
and direct N-body (the {\sl AP$^3$M} code {\sl Hydra}), tree algorithms (the
treecodes {\sl PKDGRAV} and {\sl HOT}), and hybrid tree-PM algorithms ({\sl
GADGET-2}, {\sl TPM}, and {\sl TreePM}).

The results from the code comparisons are satisfactory and not
unexpected, but also show that much more work is needed in order
to attain the required accuracy for upcoming surveys. The halo
mass function is a very stable statistic, the agreement over wide
ranges of mass being better than 5~\%. Additionally, the low mass
cutoff for individual codes can be reliably predicted by a simple
criterion.

The internal structure of halos in the outer regions of $\sim
R_{200}$ also appears to be very similar between different
simulation codes. Larger differences between the codes in the
inner region of the halos occur if the halo is not in a relaxed
state: in this case, time stepping issues might also play an
important role (e.g. particle orbit phase errors, global time
mismatches). For halos with a clear single centre, the agreement
is very good and predictions for the fall-off of the profiles from
resolution criteria hold as expected. The investigation of the
halo counts as a function of density revealed an interesting
problem with the {\sl TPM} code, the simulation suffering from a large
deficit in medium density regimes. The {\sl AMR} codes showed a large
deficit of small halos over almost the entire density regime, as
the base grid of the {\sl AMR} simulation sets a resolution limit
that is too low for the halos, as can be seen in Figure (\ref{figs:halo_comp}).

The power spectrum measurements revealed definitively more scatter
among the different codes than expected. The agreement in the
nonlinear regime is at the $5-10$~\% level, even on moderate spatial
scales around $k=10 h$~Mpc$^{-1}$. This disagreement on small
scales is connected to differences of the codes in the inner
regions of the halos. For more detailed discussion see
\citet{2007arXiv0706.1270H} and references therein.

In a detailed comparison of {\sl ENZO} and {\sl GADGET},
\citet{2005ApJS..160....1O} already pointed out that to reach
reasonable good agreement, relatively conservative criteria
for the adaptive grid refinement are needed. Furthermore, choosing a
grid resolution twice as high as the mean inter-particle distance of
the dark matter particles is recommended, to improve the small scale
accuracy of the calculation of the gravitational forces.

\section{Hydro methods}

The baryonic content of the Universe can typically be described as
an ideal fluid. Therefore, to follow the evolution of the fluid,
one usually has to solve the set of hydrodynamic equations
\begin{equation}
   \frac{\dd{\vec v}}{\dd t} = - \frac{{\vec \nabla} P}{\rho}
   -{\vec \nabla}\Phi,
\end{equation}
\begin{equation}
   \frac{\dd\rho}{\dd t} + \rho{\vec \nabla}{\vec v} = 0
\end{equation}
and
\begin{equation}
   \frac{\dd u}{\dd t} = - \frac{P}{\rho} {\vec \nabla} \cdot
   {\vec v} - \frac{\Lambda(u,\rho)}{\rho},
   \label{equation:firstlaw}
\end{equation}
which are the {\sl Euler equation}, {\sl continuity equation} and
the {\sl first law of thermodynamics}, respectively. They are
closed by an {\sl equation of state}, relating the pressure $P$ to
the internal energy (per unit mass) $u$. Assuming an ideal,
monatomic gas, this will be
\begin{equation}
   P = (\gamma -1)\rho u
\end{equation}
with $\gamma=5/3$. In the next sections, we will discuss how to
solve this set of equations, neglecting radiative losses
described by the cooling function $\Lambda(u,\rho)$;
in Sect.~4.1 we will give examples of how radiative losses or
additional sources of heat are included in cosmological codes.
We can also assume that the
${\vec \nabla}\Phi$ term will be solved using the methods described
in the previous section.

As a result of the high nonlinearity of gravitational clustering
in the Universe, there are two significant features emerging in
cosmological hydrodynamic flows; these features pose more challenges than
the typical hydrodynamic simulation without self-gravity. One
significant feature is the extremely supersonic motion around the
density peaks developed by gravitational instability, which leads
to strong shock discontinuities within complex smooth structures.
Another feature is the appearance of an enormous dynamic range in
space and time, as well as in the related gas quantities. For
instance, the hierarchical structures in the galaxy distribution
span a wide range of length scales, from the few kiloparsecs
resolved in an individual galaxy to the several tens of
megaparsecs characterising the largest coherent scale in the
Universe.

A variety of numerical schemes for solving the coupled system of
collisional baryonic matter and collisionless dark matter have
been developed in the past decades. They fall into two categories:
particle methods, which discretise mass, and grid-based methods,
which discretise space. We will briefly describe both methods in
the next two sections.

\subsection{Eulerian (grid)}

\begin{figure*}
\begin{center}
\includegraphics[width=0.95\textwidth]{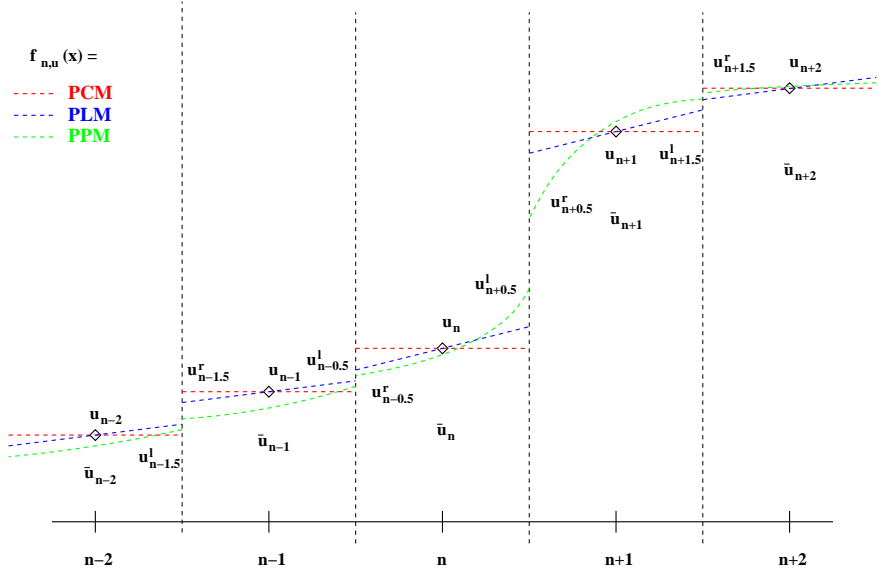}
\end{center}
\caption{Reconstruction of the principal variables ($u_n$) on the
grid using different methods, like {\sl piecewise constant}
(PCM), {\sl piecewise linear} (PLM) or {\sl piecewise parabolic} (PPM).
The reconstruction scheme then allows one to calculate cell averages
($\bar{u}_n$) as well as the left and right-hand sided values on the
cell boundaries ($u_{n\pm0.5}^l$,$u_{n\pm0.5}^r$). }
\label{figs:hydro_grid}
\end{figure*}

The set of hydrodynamical equations for an expanding Universe reads
\begin{equation}
  \frac{\partial {\vec v}}{\partial t} +
  \frac{1}{a}({\vec v}\cdot{\vec \nabla}){\vec v} +
  \frac{\dot{a}}{a}{\vec v}
  = -\frac{1}{a\rho}{\vec \nabla}P - \frac{1}{a}{\vec \nabla}\Phi,
\end{equation}
\begin{equation}
  \frac{\partial \rho}{\partial t} +
  \frac{3\dot{a}}{a}\rho +
  \frac{1}{a}{\vec \nabla}\cdot(\rho{\vec v}) = 0
\end{equation}
and
\begin{equation}
  \frac{\partial}{\partial t}(\rho u) +
  \frac{1}{a}{\vec v} \cdot {\vec \nabla}(\rho u) =
  -(\rho u + P)\left(\frac{1}{a}{\vec \nabla} \cdot {\vec v} + 3\frac{\dot{a}}{a}\right)
\end{equation}
respectively, where the right term in the last equation reflects
the expansion in addition to the usual $P\dd V$ work.

The grid-based methods solve these equations based on structured or
unstructured grids, representing the fluid. One distinguishes
{\sl primitive variables}, which determine the thermodynamic
properties, (e.g $\rho$, ${\vec v}$ or $P$) and {\sl conservative
variables} which define the conservation laws, (e.g. $\rho$,
$\rho{\vec v}$ or $\rho u$). Early attempts were
made using a central difference scheme, where fluid is only
represented by the centred cell values (e.g. {\sl central variables},
$u_n$ in Fig.~\ref{figs:hydro_grid} and derivatives are obtained by the
finite-difference representation, similar to Eq.~\ref{eq:finite1}
and \ref{eq:finite2}, see for example \citet{1992ApJS...78..341C}. Such
methods will however break down in regimes where discontinuities
appear. These methods therefore use artificial viscosity to
handle shocks (similar to the smoothed particle hydrodynamics
method described in the next section). Also, by construction, they are
only first-order accurate.

More modern approaches use reconstruction schemes, which, depending on
their order, take several neighbouring cells into account to
reconstruct the field of any hydrodynamical variable. Fig.~\ref{figs:hydro_grid}
illustrates three different reconstruction schemes, with increasing
order of accuracy, as {\sl piecewise constant method}
({\sl PCM}), {\sl piecewise linear method} \citep[{\sl PLM}, e.g.][]{1985JCoPh..59..264C}
and {\sl piecewise parabolic method} \citep[{\sl PPM},][]{1984JCoPh..54..174C}.
The shape of the reconstruction $f_{n,u}(x)$ is then used to calculate
the total integral of a quantity over the
grid cell, divided by the volume of each cell (e.g. cell average, $\hat{u}_n$),
rather than pointwise approximations at the grid centres (e.g.
central variables, $u_n$).
\begin{equation}
   \hat{u}_n = \int\limits_{x_{n-0.5}}^{x_{n+0.5}} f_{n,u}(x)
   \dd x
\label{eq:reconstruct}
\end{equation}

They are also used to calculate the left
and right-hand sided values at the cell boundaries
(e.g. $u_{n\pm0.5}^l$,$u_{n\pm0.5}^r$), which are used later as
initial conditions to solve the Riemann problem. To avoid oscillations
(e.g. the development of new extrema), additional
constraints are included in the reconstruction. For example, in the
{\sl PLM} reconstruction this is ensured by using so-called slope limiters
which estimate the maximum slope allowed for the reconstruction. One
way is to demand that the total variation among the interfaces does
not increase with time. Such so-called total variation
diminishing schemes \citep[{\sl TVD},][]{1983JCoPh..49..357H}, nowadays provide
various different slope limiters suggested by different authors.
In our example illustrated by Fig.~\ref{figs:hydro_grid}, the
so called minmod slope limiter
\begin{equation}
   \overline{\Delta u_i} = \mathrm{minmod}\left(\Theta(u_{i+1}-u_i),(u_{i+1}-u_{i-1})/2,\theta(u_i-u_{i-1})\right),
\end{equation}
where $\overline{\Delta u_i}$ is the limiter
slope within the cell $i$ and $\theta = [1,2]$, would try to fix the
slope $f'_{n-1,u}(x_{n-1})$ and $f'_{n,u}(x_n)$, such as to avoid
that $u_{n-0.5}^l$ becomes
larger than $u_{n-0.5}^r$. The so called Aldaba-type limiter
\begin{equation}
\overline{\Delta u_i} = \frac{2(u_{i+1}-u_i)(u_i-u_{i-1})+\epsilon}
                        {(u_{i+1}-u_i)^2+(u_i-u_{i-1})^2+\epsilon^2}
                   \frac{1}{2}(u_{i+1}-u_{i-1}),
\end{equation}
where $\epsilon$ is a small positive number to avoid problems in
homogeneous regions, would try to avoid
that $u_{n-0.5}^l$ is getting larger than $u_n$ and that $u_{n-0.5}^r$ is
getting smaller than $u_{n-1}$, e.g. that a monotonic profile in $u_i$
is preserved.

In the {\sl PPM} (or even higher order) reconstruction this enters as an
additional condition when finding the best-fitting polynomial function.
The additional cells which are involved in the reconstruction are often
called the {\sl stencil}. Modern, high order schemes usually have 
stencils based on at least 5 grid points and implement 
essentially non-oscillatory \citep[{\sl ENO};][]{1987JCoPh..71..231H} or
monoticity preserving ({\sl MP}) methods for reconstruction,
which maintain high-order accuracy. For every reconstruction, a {\sl
smoothness indicator} $S_n^m$ can be constructed, which is defined as the
integral over the sum of the squared derivatives of the reconstruction over
the stencil chosen, e.g.
\begin{equation}
   S_n^m = \sum_{l=1}^2 \int\limits_{x_{n-m}}^{x_{n+m}} (\Delta
   x)^{2l-1}\left(\partial_x^l f_{n,u}^m(x)\right)^2 \dd V .
\end{equation}
In the {\sl ENO} schemes, a set of candidate polynomials $p_n^m$ with
order $2 m+1$ for a set of {\sl stencils} based on different numbers
of grid cells $m$ are used to define several different reconstruction
functions $f_{n,u}^m$. Then, the reconstruction with the lowest
smoothness indicator $S_n^m$ is chosen. In this way the order of
reconstruction will be reduced around discontinuities, and oscillating
behaviour will be suppressed.

To improve on the {\sl ENO} schemes in robustness and accuracy one
can, instead of selecting the reconstruction with the best 
smoothness indicator $S_n^m$, construct the final reconstruction by
building the weighted reconstruction
\begin{equation}
   \hat{f}_u(x) = \sum_m w_m f_{n,u}^m(x),
\end{equation}
where the weights $w_m$ are a proper function of the 
smoothness indicators $S_n^m$. This procedure is not unique. \citet{Jiang_Shu__1996__WENO-scheme}
proposed defining
\begin{equation}
w_m = \frac{\alpha_m}{\sum_l\alpha_l}
\end{equation}
with
\begin{equation}
\alpha_l = \frac{C_l}{(\epsilon + S_n^l)^\beta},
\end{equation}
where $C_l$, $\epsilon$ and $\beta$ are free parameters, which for
example can be taken from \citet{1999math.....11089L}. This are the so-called
weighted essentially non-oscillatory ({\sl WENO}) schemes.
These schemes can simultaneously provide a high-order resolution for the
smooth part of the solution and a sharp, monotonic shock or contact
discontinuity transition. For a review on {\sl ENO} and {\sl WENO}
schemes, see e.g. \citet{Shu__1998__CIME__WENO}.

After the left and right-hand values at the cell boundaries
(e.g. interfaces) are reconstructed, the resulting Riemann problem
is solved, e.g. the evolution of two constant states separated by a
discontinuity. This can be done either analytically or approximately,
using left and right-handed values at the interfaces as a jump condition.

With the solution one obtains, the fluxes across these
boundaries for the time step can be calculated
and the cell averages $\hat{u}_n$
can be updated accordingly. In multiple dimensions, all these steps are
performed for each coordinate direction separately, taking only the
interface values along the individual axes into account. There are
attempts to extend the reconstruction schemes, to directly reconstruct the
principal axis of the Riemann problem in multiple dimensions, so that
then it has to be solved only once for each cell. However the
complexity of reconstructing the surface of shocks in three dimensions
has so far seen to be untraceable.

\begin{figure*}
\begin{center}
\includegraphics[width=0.6\textwidth]{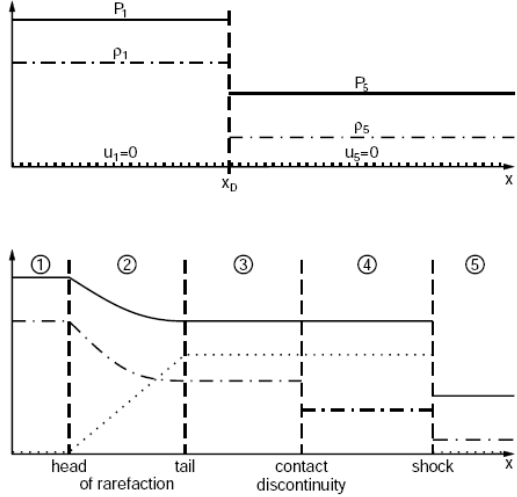}
\end{center}
\caption{The Riemann problem: The upper panel shows the initial
state, the lower panel shows the evolved problem for the case of no
relative motion between the two sides ($u_1=u_5=0$). The solid lines mark the
pressure $P$, the dashed dotted lines the density $\rho$ and the dotted
line the velocity $v$. Kindly provided by Ewald M\"uller.}
\label{figs:riemann}
\end{figure*}

How to solve the general Riemann problem, e.g. the
evolution of a discontinuity initially separating two states, can
be found in text books \citep[e.g.][]{1948sfsw.book.....C}. Here
we give only the evolution of a shock tube as an
example. This corresponds to a system where both sides are
initially at rest. Fig.~\ref{figs:riemann} shows the initial
and the evolved system. The latter can be divided into 5 regions.
The values for regions 1 and 5 are identical to the initial
configuration. Region 2 is a rarefaction wave which is determined
by the states in region 1 and 3. Therefore we are left with 6
variables to determine, namely $\rho_3$, $P_3$, $v_3$ and
$\rho_4$, $P_4$, $v_4$, where we have already eliminated the
internal energy $u_i$ in all regions, as it can be calculated from
the equation of state. As there is no mass flux through the
contact discontinuity, and as the pressure is continuous across
the contact discontinuity, we can eliminate two of the six
variables by setting $v_3=v_4=v_{\rm c}$ and $P_3=P_4=P_{\rm c}$. The general
{\sl Rankine-Hugoniot} conditions, describing the jump conditions
at a discontinuity, read
\begin{eqnarray}
  \rho_l v_l &=& \rho_r v_r \\
  \rho_l v_l^2 + P_l &=& \rho_r v_r^2 + P_r \\
  v_l \left(\rho_l(v_l^2/2+u_l) + P_l \right) &=&
  v_r \left(\rho_r(v_r^2/2+u_r) + P_r \right),
\end{eqnarray}
where we have assumed a coordinate system which moves with the
shock velocity $v_{\rm s}$. Assuming that the system is at rest in the
beginning, e.g. $v_1=v_5=0$, the {\sl first} Rankine-Hugoniot
condition for the shock between region 4 and 5 moving with a
velocity $v_{\rm s}$ (note the implied change of the coordinate system)
is in our case
\begin{equation}
   m = \rho_5 v_{\rm s} = \rho_4(v_{\rm s}-v_{\rm c})
\end{equation}
and therefore the shock velocity becomes
\begin{equation}
   v_{\rm s} = \frac{\rho_4 v_{\rm c}}{\rho_4 - \rho_5}.
\end{equation}
The {\sl second} Rankine-Hugoniot condition is
\begin{equation}
   m v_{\rm c} = \rho_4(v_{\rm s} - v_{\rm c}) v_{\rm c} = P_{\rm c} - P_5,
\end{equation}
which, combined with the first, can be written as
\begin{equation}
\rho_4\left(\frac{\rho_4v_{\rm c}}{\rho_4-\rho_5}-v_{\rm c}\right)v_{\rm c} =
P_{\rm c} -
P_5,
\end{equation}
which, slightly simplified, leads to a {\sl first condition}
\begin{equation}
(P_5-P_{\rm c})\left(\frac{1}{\rho_5}-\frac{1}{\rho_4}\right) = - v_{\rm c}^2.
\label{eq:1st}
\end{equation}
The {\sl third} Rankine-Hugoniot condition is
\begin{equation}
   m(\epsilon_4+\frac{v_{\rm c}^2}{2}-\epsilon_5) = P_{\rm c} v_{\rm c},
\end{equation}
which, by eliminating $m$, can be written as
\begin{equation}
   \epsilon_4 - \epsilon_5 = \frac{P_{\rm c}+P_5}{P_{\rm c}-P_5}\frac{v_{\rm c}^2}{2}.
\end{equation}
Using the first condition (Eq.~\ref{eq:1st}) and
assuming an ideal gas for the equation of state, one gets
\begin{equation}
  \frac{1}{\gamma-1}\left(\frac{P_{\rm
  c}}{\rho_4}-\frac{P_5}{\rho_5}\right)=\frac{P_{\rm c}+P_5}{2\rho_4\rho_5},
\end{equation}
which leads to the {\sl second condition}
\begin{equation}
   \frac{P_{\rm c}-P_5}{P_{\rm c}+P_5}=\gamma\frac{\rho_4-\rho_5}{\rho_4+\rho_5}.
\label{eq:2nd}
\end{equation}
The {\sl third condition} comes from the fact that the entropy
($\propto \ln (P/\rho^\gamma)$ stays constant in the
rarefaction wave, and therefore one can write it as
\begin{equation}
   \frac{P_1}{P_{\rm c}}=\left(\frac{\rho_1}{\rho_3}\right)^\gamma.
\label{eq:3rd}
\end{equation}
The {\sl fourth condition} comes from the fact that the Riemann
Invariant
\begin{equation}
  v + \int \frac{c}{\rho} \dd \rho
\end{equation}
is a constant, which means that
\begin{equation}
  v_{\rm c} + \int \frac{c_3}{\rho_3} \dd \rho = v_1 + \int
  \frac{c_1}{\rho_1} \dd \rho,
\end{equation}
where $c=\sqrt{\gamma P/\rho}$ denotes the sound velocity, with which
the integral can be written as
\begin{equation}
    \int \frac{c}{\rho} \dd \rho =
    \frac{2}{\gamma-1}\sqrt{\frac{\gamma P}{\rho}}.
\end{equation}
Therefore, the fourth condition can be written as
\begin{equation}
  v_{\rm c}+ \frac{2}{\gamma-1}\sqrt{\frac{\gamma P_{\rm c}}{\rho_3}} =
  \frac{2}{\gamma-1}\sqrt{\frac{\gamma P_1}{\rho_1}}.
\label{eq:4th}
\end{equation}
Combining all 4 conditions (equations
\ref{eq:1st}, \ref{eq:2nd}, \ref{eq:3rd} and \ref{eq:4th}) and defining
the initial density ratio $\lambda=\rho_1/\rho_5$ one gets the non
linear, algebraic equation
\begin{equation}
\frac{\rho_1}{\rho_5}\frac{1}{\lambda}\frac{(1-P)^2}{\gamma(1+P)-1+P}
=
\frac{2\gamma}{(\gamma-1)^2}\left[1-\left(\frac{P}{\lambda}\right)^{(\gamma-1)/(2\gamma)}\right]^2
\end{equation}
for the pressure ratio $P=P_{\rm c}/P_5$. Once $P_{\rm c}$ is known from solving
this equation, the remaining unknowns can be inferred step by step from
the four conditions.

There are various approximate methods to solve the Riemann problem, including the
so-called {\sl ROE method} \citep[e.g.][]{1999JCoPh.154..284P}, {\sl HLL/HLLE}
method \citep[e.g. see][]{Harten:1983:UDG,Einfeldt:1988:GTM,1991JCoPh..92..273E} 
and {\sl HLLC} \citep[e.g. see][]{2005JCoPh.203..344L}. A description of all
these methods is outside the scope of this review, so we redirect the
reader to the references given or textbooks like \citet{L2002}.

At the end of each time step, one has to compute the updated
central values $u_n$ from the updated  cell average
values $\hat{u}_n$. Normally, this would imply inverting equation
(\ref{eq:reconstruct}), which is not trivial in the general case.
Therefore, usually an additional constraint is placed on the
reconstruction method, namely that the
reconstruction fulfills $u_n = \hat{u}_n$. In this case the last step
is trivial.

In general, the grid-based methods suffer from limited spatial
resolution, but they work extremely well in both low- and
high-density regions, as well as in shocks. In cosmological
simulations, accretion flows with large Mach numbers (e.g. $M > 100$)
are very common. Here, following the total energy in the
hydrodynamical equations, one can get inaccurate thermal energy, leading
to negative pressure, due to discretisation errors when the kinetic
energy dominates the total energy. In such cases, as suggested by
\citet{1993ApJ...414....1R} and \citet{1995CoPhC..89..149B}, the numerical schemes usually
switch from formulations solving the total energy to formulations
based on solving the internal energy in these hypersonic flow regions.

In the cosmological
setting, there are the {\sl TVD}-based codes including those of \citet{1993ApJ...414....1R}
and \citet{2006astro.ph.11863L} ({\sl CosmoMHD}), the moving-mesh scheme
\citep{1998ApJS..115...19P} and the {\sl PLM}-based code {\sl ART}
\citep{1997ApJS..111...73K,2002APS..APR.E2004K}. 
The {\sl PPM}-based codes include those of \citet{1992ApJS...80..753S}
({\sl Zeus}), \citet{1995CoPhC..89..149B} ({\sl ENZO}),
\citet{2000ApJ...536..122R} ({\sl COSMOS}) and
\citet{2000ApJS..131..273F} ({\sl FLASH}). There is also the {\sl WENO}-based
code by \citet{2004ApJ...612....1F}.

\subsection{Langrangian (SPH)}

The particle methods include variants of smoothed particle
hydrodynamics \citep[{\sl SPH};][]{1977MNRAS.181..375G,1977AJ.....82.1013L} such as
those of \citet{1988MNRAS.235..911E}, \citet{1989ApJS...70..419H}, 
\citet{1993MNRAS.265..271N}, \citet{1995ApJ...452..797C} ({\sl
Hydra}), \citet{1996MNRAS.278.1005S} ({\sl GRAPESPH}), 
\citet{1998ApJS..116..155O}, and \citet{springel2001,springel2005} 
({\sl GADGET}). The {\sl SPH} method
solves the Lagrangian form of the Euler equations and can
achieve good spatial resolutions in high-density regions, but it
works poorly in low-density regions. It also suffers from degraded
resolution in shocked regions due to the introduction of a sizable
artificial viscosity. \citet{2007MNRAS.380..963A} argued that whilst
Eulerian grid-based methods are able to resolve and treat dynamical
instabilities, such as {\sl Kelvin-Helmholtz} or {\sl Rayleigh-Taylor},
these processes are poorly resolved by existing {\sl SPH} techniques.
The reason for this is that {\sl SPH}, at least in its standard
implementation, introduces spurious pressure forces on particles
in regions where there are steep density gradients, in particular near
contact discontinuities.
This results in a boundary gap of the size of an {\sl SPH} smoothing kernel
radius, over which interactions are severely damped.
Nevertheless, in the cosmological context, the adaptive nature of the
{\sl SPH} method compensates for such shortcomings, thus making
{\sl SPH} the most commonly used method in numerical
hydrodynamical cosmology.

\subsubsection{Basics of SPH}

The basic idea of {\sl SPH} is to discretise the fluid by mass elements
(e.g. particles), rather than by volume elements as in the Eulerian
methods. Therefore it is immediately clear that
the mean inter-particle distance in collapsed objects will be smaller
than in underdense regions;
the scheme will thus be adaptive in spatial resolution by
keeping the mass resolution fixed. For a comprehensive review see
\citet{1992ARA&A..30..543M}.
To build continuous fluid
quantities, one starts with a general definition of a kernel smoothing
method
\begin{equation}
   \left< A({\vec x}) \right> = \int W({\vec x} - {\vec x}',h) A({\vec x}') \dd {\vec x}',
\end{equation}
which requires that the kernel is normalised (i.e. $\int W({\vec
x},h) \dd {\vec x} = 1$) and collapses to a delta function if the
smoothing length $h$ approaches zero, namely $W({\vec x},h)\rightarrow\delta({\vec x})$ for $h \rightarrow 0$.

One can write down the continuous fluid quantities (e.g. $\left<A({\vec x})\right>$)
based on the discretised values $A_j$ represented by the set of the individual
particles $m_j$ at the position ${\vec x}_j$ as
\begin{equation}
   \left<A_i\right> = \left< A({\vec x}_i) \right> = \sum_j
   \frac{m_j}{\rho_j} A_j W({\vec x_i} - {\vec x_j},h) \, ,
   \label{SPH:base}
\end{equation}
where we assume that the kernel depends only on
the distance modulus (i.e. $W(|{\vec x} - {\vec x}'|,h)$) and we
replace the volume element of the integration, $\dd {\vec x} =
\dd^3 x$, with the ratio of the mass and density $m_j/\rho_j$ of the particles. Although this
equation holds for any position ${\vec x}$ in space, here we are
only interested in the fluid representation at the original
particle positions ${\vec x}_i$, which are the only locations where
we will need the fluid representation later on. It is important to
note that for kernels with compact support (i.e. $W({\vec
x},h)=0$ for $|{\vec x}|>h$) the summation does not have to be done over
all the particles, but only over the particles within the sphere of
radius $h$, namely the neighbours around the particle $i$ under
consideration. Traditionally, the most frequently used kernel is the
$B_2$-spline, which can be written as
\begin{equation}
   W(x,h)=\frac{\sigma}{h^\nu}\left\{\begin{array}{ll}
      1 - 6 \left(\frac{x}{h}\right)^2 + 6 \left(\frac{x}{h}\right)^3 \;\;& 0 \le \frac{x}{h} < 0.5 \\
      2 \left(1 - \frac{x}{h}\right)^3                              & 0.5 \le \frac{x}{h} < 1 \\
      0                                                             & 1 \le \frac{x}{h} \\
   \end{array} \right. , \label{SPH:kern}
\end{equation}
where $\nu$ is the dimensionality (e.g. 1, 2 or 3) and $\sigma$ is the
normalisation
\begin{equation}
   \sigma=\left\{ \begin{array}{ll}
      \frac{16}{3}       & \nu=1 \\
      \frac{80}{7\pi} \;\;  & \nu=2 \\
      \frac{8}{\pi}     & \nu=3 \; . \\
   \end{array} \right.
\end{equation}
Sometimes, spline kernels of higher order are used for very special
applications; however the $B_2$ spline kernel turns out to be the
optimal choice in most cases.

When one identifies $A_i$ with the density $\rho_i$,
$\rho_i$ cancels out on the right hand side of Eq.~\ref{SPH:base}, 
and we are left with the density estimate
\begin{equation}
   \left<\rho_i\right> = \sum_j m_j W({\vec x_i} - {\vec x_j},h),
   \label{SPH:density}
\end{equation}
which we can interpret as the density of the fluid element
represented by the particle $i$.

Now even derivatives can be calculated as
\begin{equation}
   {\vec \nabla} \left<A_i\right> = \sum_j \frac{m_j}{\rho_j} A_j {\vec \nabla}_i W({\vec x_i} - {\vec x_j},h),
   \label{SPH:derivative}
\end{equation}
where ${\vec \nabla}_i$ denotes the
derivative with respect to ${\vec x}_i$. A pairwise symmetric
formulation of derivatives in {\sl SPH} can be obtained by making use of
the identity
\begin{equation}
(\rho{\vec\nabla})\cdot{}A={\vec\nabla}(\rho\cdot{}A)-\rho\cdot({\vec\nabla}A),
\end{equation}
which allows one to re-write a derivative as
\begin{equation}
   {\vec \nabla} \left<A_i\right> = \frac{1}{\rho_i} \sum_j
   m_j (A_j-A_i) {\vec \nabla}_i W({\vec x_i} - {\vec x_j},h).
   \label{SPH:symmetric_derivative}
\end{equation}
Another way of symmetrising the derivative is to use the identity
\begin{equation}
   \frac{{\vec \nabla} A}{\rho} = {\vec
   \nabla}\left(\frac{A}{\rho}\right) + \frac{A}{\rho^2}{\vec \nabla}\rho,
\end{equation}
which then leads to the following form of the derivative:
\begin{equation}
   {\vec \nabla} \left<A_i\right> = \rho_i \sum_j
   m_j \left(\frac{A_j}{\rho_j^2} + \frac{A_i}{\rho_i^2}\right) {\vec \nabla}_i W({\vec x_i} - {\vec x_j},h).
   \label{SPH:quadratic_derivative}
\end{equation}

\subsubsection{The fluid equations}

By making use of these identities, the Euler equation can be written as
\begin{equation}
   \frac{\dd {\vec v}_i}{\dd t} = - \sum_j
   m_j \left(\frac{P_j}{\rho_j^2} + \frac{P_i}{\rho_i^2} + \Pi_{ij}\right) {\vec \nabla}_i W({\vec x_i} - {\vec x_j},h).
   \label{SPH:euler}
\end{equation}
By combining the above identities and averaging the result, the term
$-(P/\rho){\vec \nabla}\cdot{\vec v}$ from the first law of
thermodynamics can similarly be written as
\begin{equation}
  \frac{\dd u_i}{\dd t} = \frac{1}{2} \sum_j
   m_j \left(\frac{P_j}{\rho_j^2} + \frac{P_i}{\rho_i^2} +
   \Pi_{ij}\right) \left({\vec v}_j - {\vec v}_i\right){\vec \nabla}_i W({\vec x_i} - {\vec x_j},h).
   \label{SPH:firstlaw}
\end{equation}
Here we have added a term $\Pi_{ij}$ which is the so-called
{\sl artificial viscosity}. This term is usually needed to capture
shocks and its construction is  similar to other hydro-dynamical
schemes. Usually, one adopts the form proposed by
\citet{1983JCoPh..52....374S} and \citet{1995JCoPh.121..357B},
which includes a bulk viscosity and a von~Neumann-Richtmeyer
viscosity term, supplemented by a term controlling
angular momentum transport in the presence of shear flows at low
particle numbers  \citep{1996IAUS..171..259S}. Modern schemes
implement a form of the artificial viscosity as proposed by
\cite{1997JCoPh..136....298S} based on an analogy with Riemann
solutions of compressible gas dynamics. To reduce this artificial
viscosity, at least in those parts of the flows where there are no shocks,
one can follow the idea proposed by \cite{1997JCoPh..136....41S}:
every particle carries its own artificial viscosity, which
eventually decays outside the regions which undergo shocks. A detailed
study of the implications on the ICM of such an implementation
can be found in \citet{2005MNRAS.364..753D}.

The {\sl continuity equation} does not have to be evolved explicitly, as
it is automatically fulfilled in Lagrangian methods. As shown
earlier, density is no longer a variable but can be, at any point,
calculated from the particle positions. Obviously, mass
conservation is guaranteed, unlike volume conservation: in other words,
the sum of the volume elements associated with all of the particles might
vary with time, especially when strong density
gradients are present.

\subsubsection{Variable smoothing length}

Usually, the smoothing length $h$ will be allowed to vary for each
individual particle $i$ and is determined by finding the radius $h_i$ of a
sphere which contains $n$ neighbours. Typically,
different numbers $n$ of neighbours are chosen by different authors,
ranging from 32 to 80. In principle, depending on the kernel,
there is an optimal choice of neighbours (e.g. see
\citealt{1986desd.book.....S} or similar books). 
However, one has to find a compromise
between a large number of neighbours, leading to larger systematics
but lower noise in the density estimates (especially in regions with large density
gradients) and a small number of neighbours, leading to larger sample
variances for the density estimation. In general, once every
particle has its own smoothing length, a symmetric kernel $W({\vec
x}_i - {\vec x}_j,h_i,h_j) = {\bar W}_{ij}$ has to be constructed
to keep the conservative form of the formulations of the
hydrodynamical equations. There are two main variants used in
the literature: one is the kernel average ${\bar W}_{ij} = (W({\vec
x}_i - {\vec x}_j,h_i)+W({\vec x}_i - {\vec x}_j,h_j))/2$, the
other is an average of the smoothing length ${\bar W}_{ij} = W({\vec
x}_i - {\vec x}_j,(h_i+h_j)/2)$. The former
is the most commonly used approach.

Note that in all of the derivatives discussed above, it is
assumed that $h$ does not depend on the position ${\vec x}_j$.
Thus, by allowing the
smoothing length $h_i$ to be variable for each particle, one
formally neglects the correction term $\partial W/\partial h$, which
would appear in all the derivatives. In general, this correction term cannot be
computed trivially and therefore many implementations do not take it
into account. It is well known that such formulations are poor at conserving numerically
both internal energy and entropy at the same time,
independently of the use of internal energy or entropy in
the formulation of the first law of thermodynamics, see
\citet{1993ApJ...404..717H}. In the
next subsection, we present a way of deriving the equations which
include these correction terms $\partial W/\partial h$; this equation
set represents a formulation which conserves numerically both entropy and
internal energy.

\subsubsection{The entropy conservation formalism}

To derive a better formulation of the {\sl SPH} method,
\citet{springel02} started from the entropic function
$A=P/\rho^\gamma$, which will be conserved in adiabatic flows. The
internal energy per unit mass can be inferred from this entropic
function as
\begin{equation}
u_i = \frac{A_i}{\gamma-1}\rho_i^{\gamma-1}
\end{equation}
at any time, if needed. Entropy will be generated by shocks, which
are captured by the artificial viscosity $\Pi_{ij}$ and therefore
the entropic function will evolve as
\begin{equation}
  \frac{\dd A_i}{\dd t} = \frac{1}{2} \frac{\gamma-1}{\rho_i^{\gamma-1}} \sum_j
   m_j \Pi_{ij} \left({\vec v}_j - {\vec v}_i\right){\vec \nabla}_i
{\bar W}_{ij}.
\end{equation}
The Euler equation can be derived starting by defining the
Lagrangian of the fluid as
\begin{equation}
   L({\vec q},\dot{\vec q}) = \frac{1}{2}\sum_i m_i \dot{\vec
   x}_i^2 - \frac{1}{\gamma-1}\sum_i m_i A_i \rho_i^{\gamma-1}
\end{equation}
which represents the entire fluid and has the coordinates ${\vec q} = ({\vec x}_1, ... ,{\vec x}_N,h_1,
..., h_N)$. The next important step is to define constraints, which
allow an unambiguous association of $h_i$ for a chosen
number of neighbours $n$. This can be done by requiring that the
kernel volume contains a constant mass for the estimated density,
\begin{equation}
\phi_i({\vec q}) = \frac{4\pi}{3}h_i^3\rho_i - nm_i = 0.
\end{equation}
The equation of motion can be obtained as the solution of
\begin{equation}
  \frac{\dd}{\dd t}\frac{\partial L}{\partial\dot{\vec q}_i} -
  \frac{\partial L}{\partial {\vec q}_i} = \sum_j \lambda_j
  \frac{\partial \phi_j}{\partial {\vec q}_i},
\end{equation}
which - as demonstrated by \citet{springel02} - can be written as
\begin{equation}
   \frac{\dd {\vec v}_i}{\dd t} = - \sum_j
   m_j \left(f_j\frac{P_j}{\rho_j^2}{\vec \nabla}_i W({\vec x_i} - {\vec x_j},h_j) +
             f_i\frac{P_i}{\rho_i^2}{\vec \nabla}_i W({\vec x_i} - {\vec x_j},h_i) +
             \Pi_{ij} {\vec \nabla}_i {\bar W}_{ij} \right),
\end{equation}
where we already have included the additional term due to the
artificial viscosity $\Pi_{ij}$, which is needed to capture shocks.
The coefficients $f_i$ incorporate fully the variable smoothing length
correction term and are defined as
\begin{equation}
f_i = \left(1 + \frac{h_i}{3\rho_i} \frac{\partial
\rho_i}{\partial h_i}\right)^{-1}.
\end{equation}
Note that in addition to the correction terms, which can be easily
calculated together with the density estimation, this formalism also avoids
all the ambiguities we saw in the derivations of the equations in the
previous section. This formalism defines how the kernel
averages (symmetrisation) have to be taken, and also fixes $h_i$ to
unambiguous values. For a detailed derivation of this formalism and
its conserving capabilities see \citet{springel02}.

\subsection{Code Comparison}

\begin{figure*}
\begin{center}
\includegraphics[width=0.49\textwidth]{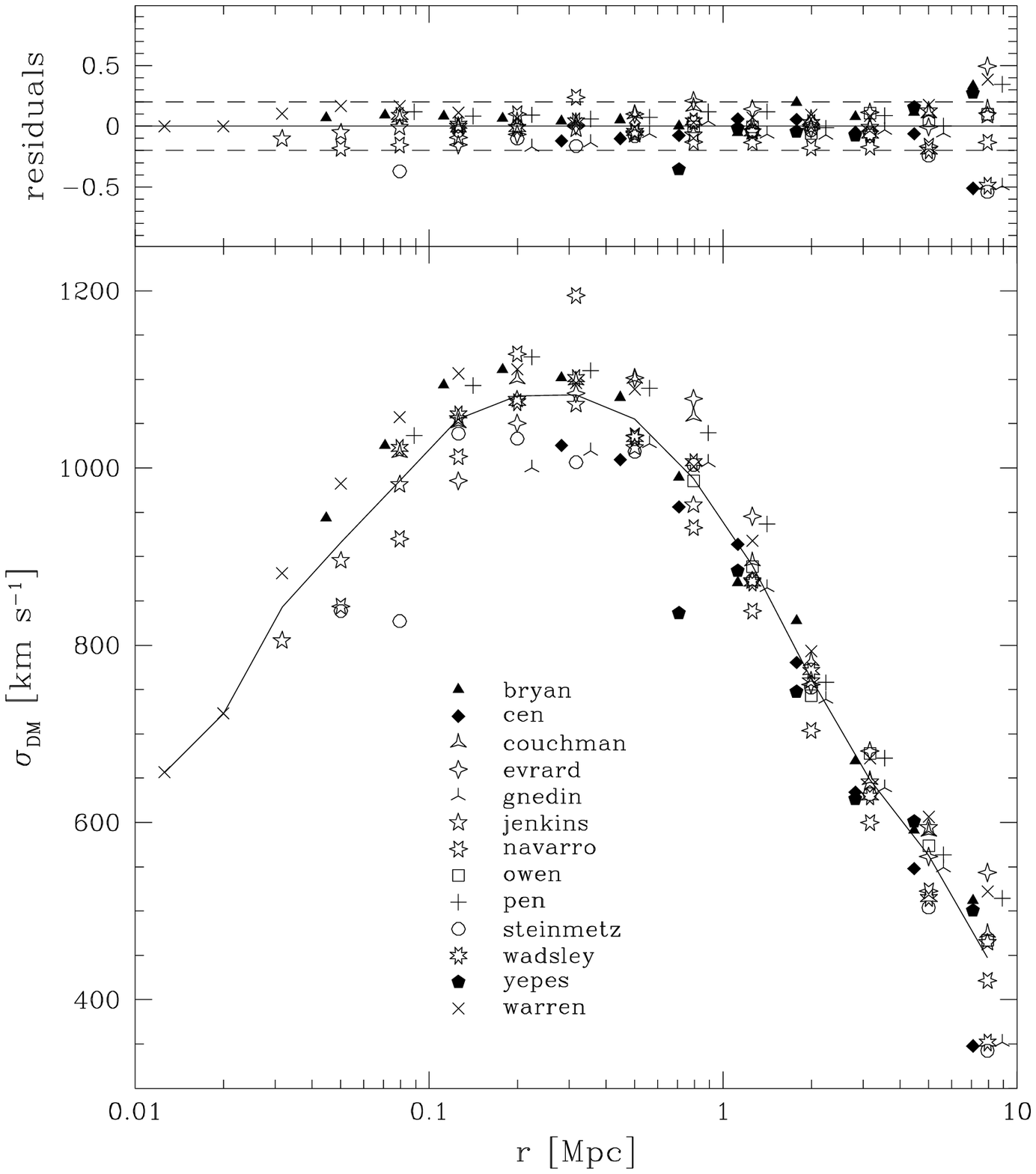}
\includegraphics[width=0.49\textwidth]{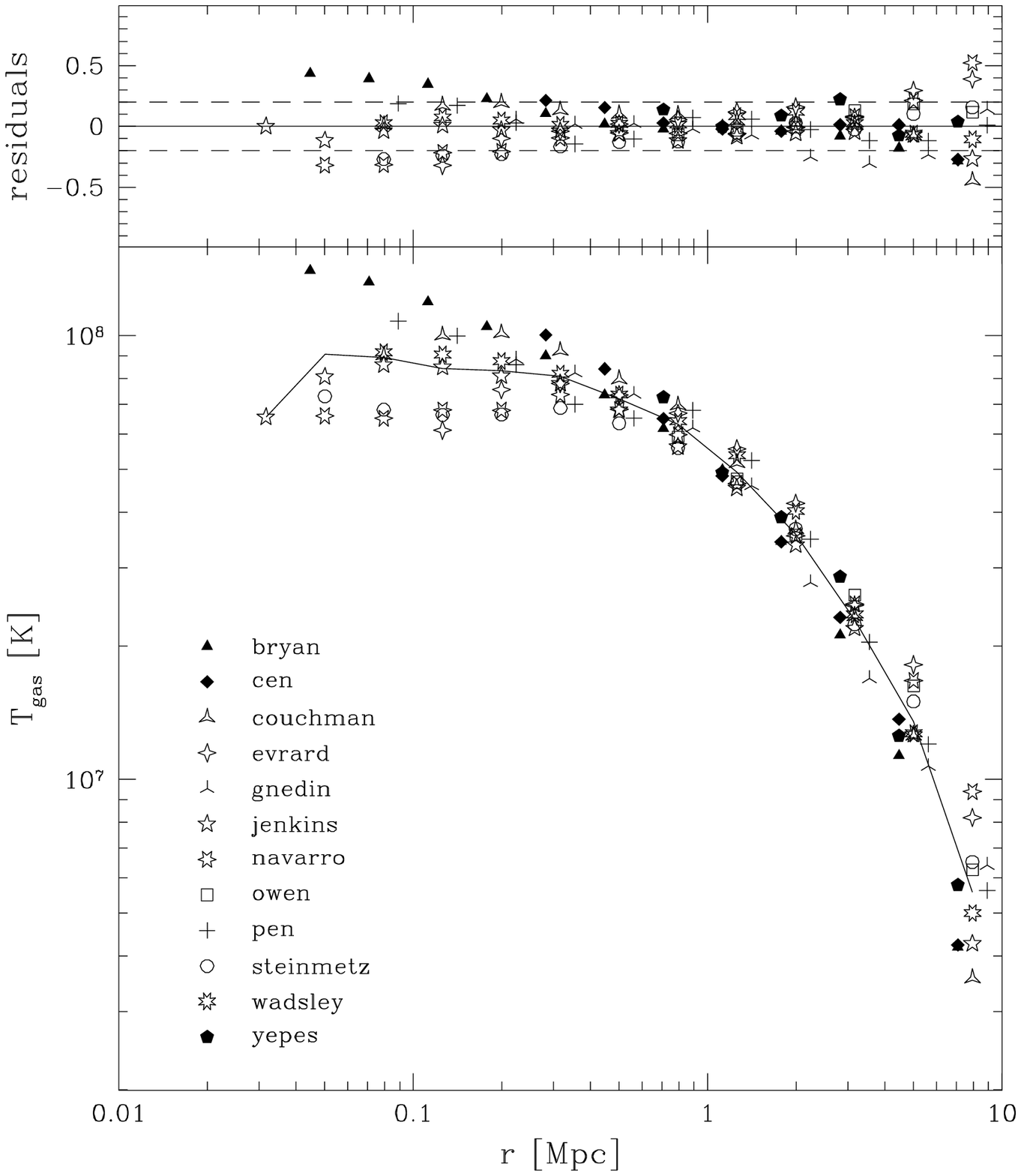}
\end{center}
\caption{One-dimensional velocity dispersion profile (left panel)
and gas temperature profile (right panel) of the cluster at $z=0$ of the Santa Barbara
Comparison Project \citep{1999ApJ...525..554F}. The solid line is the profile averaged
over the 12 simulations. The symbols correspond to individual simulations. The
crosses in the left panel correspond to a dark-matter only simulation. The top
panels show the residual from the mean profile. Taken from \citet{1999ApJ...525..554F}.}
\label{figs:sbarbara}
\end{figure*}

The Eulerian and Lagrangian approaches described in the previous sections
should provide the same results when applied to the same problem, like
the interaction of multi-phase fluids \citep{2007MNRAS.380..963A}.
To verify that the code correctly solves the hydrodynamical set of equations,
each code is usually tested against problems whose solution is known
analytically. In practice, these are shock tubes or spherical collapse
problems. In cosmology, a relevant test is to compare the results
provided by the codes when they simulate the formation of cosmic structure,
when finding an analytic solution is impractical; for example
\citet{2005ApJS..160....1O} compare
the thermodynamical properties of the intergalactic medium predicted
by the {\sl GADGET} ({\sl SPH}-based) and {\sl ENZO} (grid-based) codes. Another
example of a comparison between grid-based and {\sl SPH}-based codes can be
found in \citet{1994ApJ...430...83K}.

A detailed comparison of hydrodynamical codes which simulate the
formation and evolution of a cluster was provided by the Santa Barbara
Cluster Comparison Project \citep{1999ApJ...525..554F}. \citet{1999ApJ...525..554F}
comprised 12 different groups, each using
a code either based on the {\sl SPH} technique (7 groups) or
on the grid technique (5 groups).
Each simulation started
with identical initial conditions of an individual massive cluster in a flat
CDM model with zero cosmological constant.
Each group was free to decide resolution, boundary
conditions and the other free parameters of their code.
The simulations were performed ignoring radiative losses and the simulated clusters
were compared at $z=0.5$ and $z=0$.

The resulting dark matter properties were similar: it was found
a 20~\% scatter
around the mean density and velocity dispersion profiles (left panel
of Fig.~\ref{figs:sbarbara}).
A similar agreement was also obtained for many of the gas
properties, like the temperature profile (right panel of Fig.~\ref{figs:sbarbara}) or the ratio
of the specific dark matter kinetic energy and the
gas thermal energy.

Somewhat larger differences are present for the inner part of the
temperature or entropy profiles and more recent implementations
have not yet cured this problem.
The largest discrepancy was in the total X-ray luminosity.
This quantity is proportional to the square of the gas density,
and resolving the cluster central region within the core
radius is crucial: the simulations resolving this region
had a spread of 2.6 in the total X-ray luminosity, compared to a
spread of 10 when all the simulations were included.
\citet{1999ApJ...525..554F} also concluded that a large
fraction of the discrepancy, when excluding the X-ray luminosity result,
was due to differences in the internal
timing of the simulations: these differences produce artificial time shifts
between the outputs of the various simulations even if the outputs
are formally at the same cosmic time. This reflects mainly the
underlying dark matter treatment, including chosen force accuracy,
different integration schemes and choice of time steps used, as
described in the previous sections.
A more worrisome difference between the different codes is the
predicted baryon fraction and its profile within the cluster. Here
modern schemes still show differences \citep[e.g. see][]
{2006MNRAS.365.1021E,2005ApJ...625..588K}, which 
makes it difficult to use simulations to calibrate the systematics in
the cosmological test based on the cluster baryon fraction.

To date, the comparisons described in the literature show a
satisfactory agreement between the two approaches, with residual
discrepancies originating from the known weaknesses which are specific
to each scheme. A further
limitation of these comparisons is that, in most cases, the
simulations are non-radiative. However, at the current state of the
art, performing comparisons of simulations including radiative losses
is not expected to provide robust results. As described in the next
section, the first relevant process that needs to be added is
radiative cooling: however, depending on the square of the gas
density, cooling increases with resolution without any indication of
convergence, see for example Fig.~13, taken from
\citet{2006MNRAS.367.1641B}.  At the next level of complexity, star
formation and supernova feedback occur in regions which have a size
many orders of magnitude smaller than the spatial resolution of the
cosmological simulations. Thus, simulations use phenomenological
recipes to describe these processes, and any comparison would largely
test the agreement between these recipes rather than identify the
inadequacy of the numerical integration schemes.

\section{Adding complexity}

In this section, we will give a brief overview of how astrophysical processes,
that go beyond the description of the gravitational instability and
of the hydrodynamical flows are usually included in simulation
codes.

\subsection{Cooling}

\begin{figure*}
\begin{center}
\includegraphics[width=0.7\textwidth]{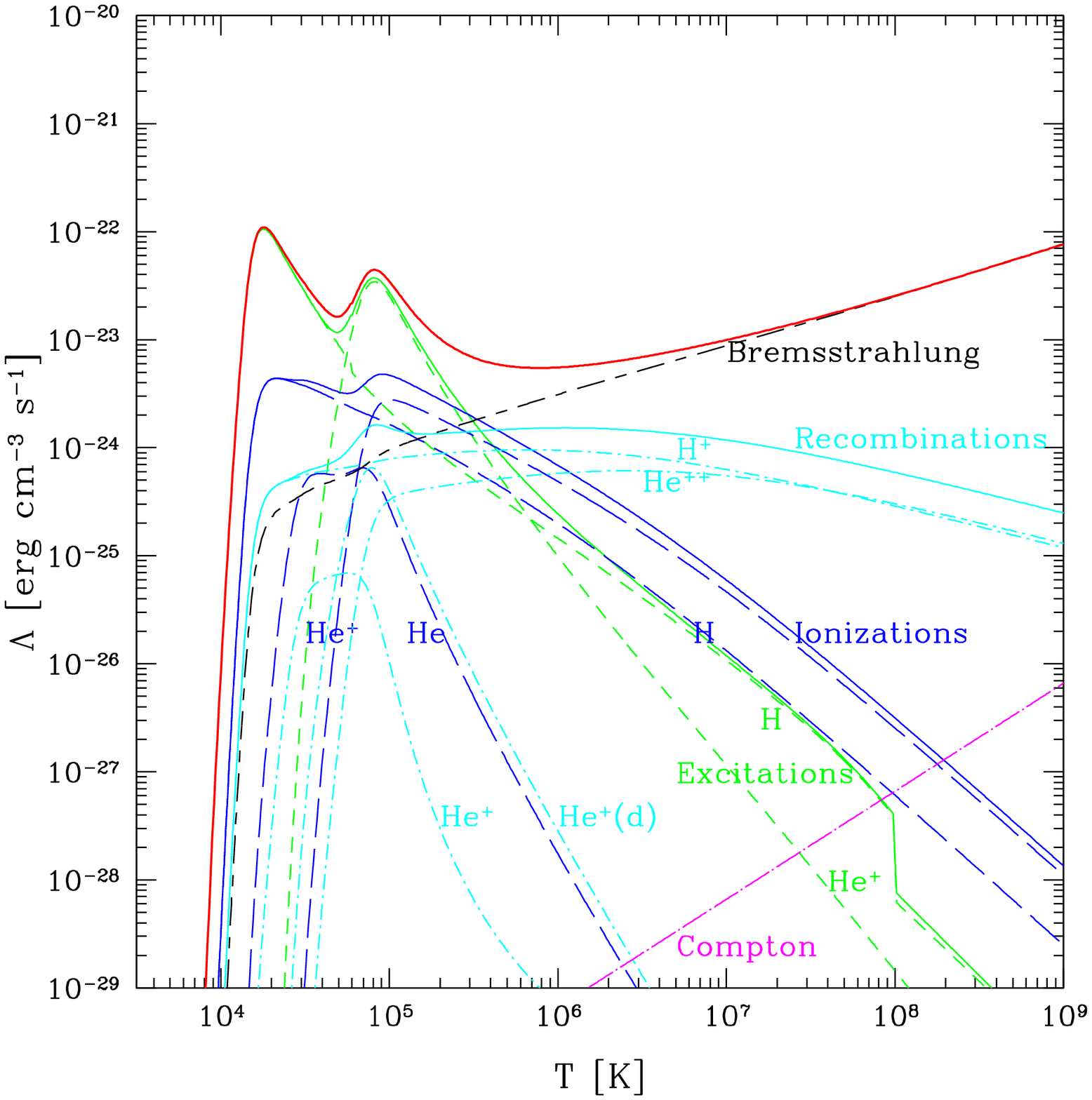}\\
\includegraphics[width=0.7\textwidth]{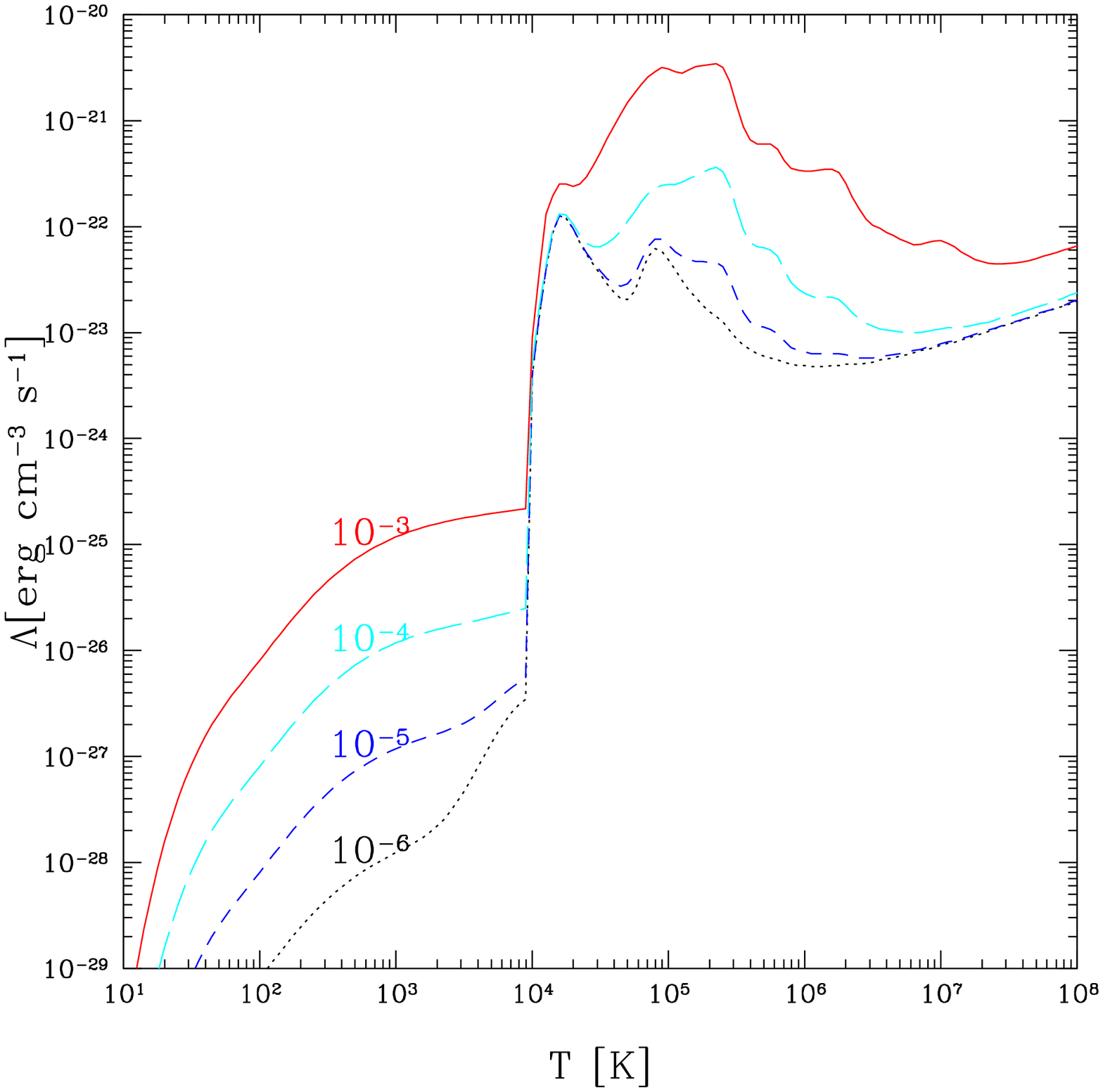}
\end{center}
\caption{The top panel shows the total cooling curve (solid
line) and its composition from different processes for a
primordial mixture of H and He. The bottom panel shows how the total
cooling curve will change as a function of different metallicity,
as indicated in the plot (in absolute values). The part below $10^4$~K also
takes into account cooling by molecules (e.g. HD and H$_2$) and metal lines.
Taken from \protect\citet{2007MNRAS.379..963M}.}
\label{figs:cooling}
\end{figure*}

We discuss here how the $\Lambda(u,\rho)$ term is usually added in the
first law of thermodynamics, described by Eq.~\ref{equation:firstlaw}, 
and its consequences.

In cosmological applications, one is usually interested in structures
with virial temperatures larger than $10^4$~K. In standard
implementations of the cooling function $\Lambda(u,\rho)$, one assumes
that the gas is optically thin and in ionisation equilibrium. It is
also usually assumed that three-body cooling processes are
unimportant, so as to restrict the treatment to two-body
processes. For a plasma with primordial composition of H and He, these
processes are collisional excitation of \ion{H}{i} and \ion{He}{ii}, 
collisional
ionisation of \ion{H}{i}, \ion{He}{i} and \ion{He}{ii}, 
standard recombination of
\ion{H}{ii}, \ion{He}{ii} and \ion{He}{iii}, 
dielectric recombination of \ion{He}{ii}, and
free-free emission (Bremsstrahlung).  The collisional ionisation and
recombination rates depend only on temperature. Therefore, in the absence
of ionising background radiation one can solve the resulting rate
equation analytically. This leads to a cooling function
$\Lambda(u)/\rho^2$ as illustrated in the top panel of Fig.~\ref{figs:cooling}.
In the presence of ionising background
radiation, the rate equations can be solved iteratively. Note that for
a typical cosmological radiation background \citep[e.g. UV background from
quasars, see][]{1996ApJ...461...20H}, the shape of the cooling function
can be significantly altered, especially at low densities. For a more
detailed discussion see for example \citet{1996ApJS..105...19K}.
Additionally, the presence of metals will drastically increase the
possible processes by which the gas can cool. As it becomes computationally
very demanding to calculate the cooling function in this case, one usually resorts
to a pre-computed, tabulated cooling function. As an example, the
bottom panel of Fig.~\ref{figs:cooling}, at temperatures above
$10^5$~K, shows the tabulated cooling function by
\citet{1993ApJS...88..253S} for different metallicities of the gas,
keeping the ratios of the different metal species fixed to solar
values. Note that almost all implementations solve the above rate
equations (and therefore the cooling of the gas) as a ``sub time step''
problem, decoupled from the hydrodynamical treatment. In practice this
means that one assumes that the density is fixed across the time step.
Furthermore, the time step of the underlying hydrodynamical simulation are 
in general, for practical reasons, not controlled by or related to the cooling
time-scale. The resulting uncertainties introduced by these
approximations have not yet been deeply explored and 
clearly leave room for future investigations.

For the formation of the first objects in haloes with virial
temperatures below $10^4$~K, the assumption of ionisation equilibrium
no longer holds. In this case, one has to follow the
non-equilibrium reactions, solving the balance equations for the
individual levels of each species during the cosmological
evolution. In the absence of metals, the main coolants are H$_2$
and H$_2^+$ molecules \citep[see][]{1997NewA....2..181A}. HD molecules can also
play a significant role. When metals are present, many more reactions
are available and some of these can contribute significantly to the
cooling function below $10^4$~K. This effect is clearly visible in the
bottom panel of Fig.~\ref{figs:cooling}, for $T<10^4$~K. For more
details see \citet{1998A&A...335..403G} or \citet{2007MNRAS.379..963M}
and references therein.

\subsection{Star formation and feedback}

\begin{figure*}
\begin{center}
\includegraphics[width=0.7\textwidth]{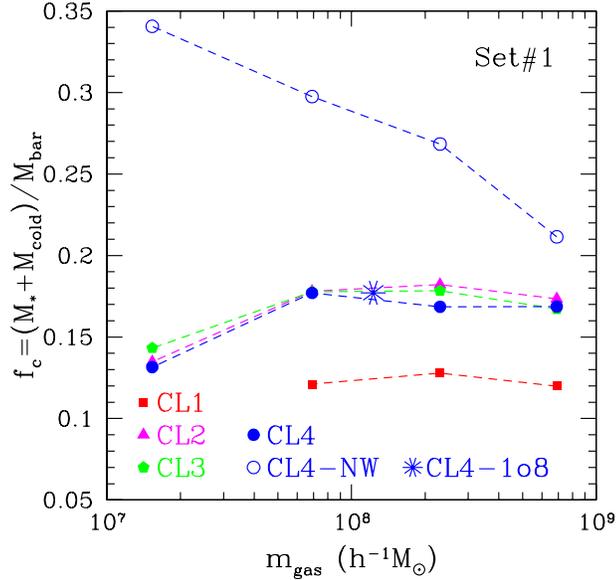}
\end{center}
\caption{The fraction of cooled baryons $f_{\rm c}$ as a function of the mass of
the gas particle, for 4 different clusters at different
resolutions is shown. Filled symbols are for the runs including
kinetic feedback (e.g. winds), the open circles are re-simulations of
one of the clusters with wind feedback turned off. The asterisk is for
one of the clusters run at very high resolution
using fewer, but 8 times heavier, gas particles than normal, so that the gas particle
mass is similar to that of the DM particles in the high--resolution
region. Taken from \citet{2006MNRAS.367.1641B}.}
\label{figs:coldbar}
\end{figure*}

Including radiative losses in simulations causes two numerical
problems. Firstly, cooling is a runaway process and, at the typical
densities reached at the centres of galaxy clusters, the cooling time
becomes significantly shorter than the Hubble time. As a consequence, a
large fraction of the baryonic component can cool down and condense
out of the hot phase. Secondly, since cooling is proportional to the
square of the gas density, its efficiency is quite sensitive to the
presence of the first collapsing small halos, where cooling takes
place, and therefore on numerical resolution.

To deal with these issues, one has to include in the code a suitable
recipe to convert the reservoir of cold and dense gas into
collisionless stars. Furthermore, this stellar component should
represent the energy feedback from supernova explosions, which
ideally would heat the cold gas, so as to counteract the cooling
catastrophe.

As for star formation, a relatively simple recipe is that originally
introduced by
\cite{1996ApJS..105...19K}, which is often used in cosmological
simulations. According to this prescription, for a gas particle to be
eligible to form stars, it must have a convergent flow,
\begin{equation}
  {\vec \nabla} {\vec v_i} < 0\,,
\end{equation}
and have density in excess of some threshold value, e.g.
\begin{equation}
   \rho_i > 0.1\ {\rm atoms\,cm}^{-3}.
\label{eqn:rho_tresh}
\end{equation}
These criteria are complemented by requiring the gas to be Jeans
unstable, that is
\begin{equation}
\frac{h_i}{c_i} > \frac{1}{\sqrt{4\pi {\rm G} \rho_i}}\,,
\end{equation}
where $h_i$ is either the {\sl SPH} smoothing length or the mesh size
for Eulerian codes and $c_i$ is the local sound speed. This
indicates that the individual resolution element gets gravitationally
unstable. At high redshift, the physical
density can easily exceed the threshold given in Eq.~\ref{eqn:rho_tresh}, 
even for particles not belonging to virialised
halos. Therefore one usually applies a further condition on the gas
overdensity,
\begin{equation}
\frac{\rho_i}{\rho_{\rm mean}} > 55.7,
\end{equation}
which restricts star formation to collapsed, virialised regions. Note
that the density criterion is the most important one. Particles
fulfilling it in almost all cases also fulfill the other two criteria.

Once a gas particle is eligible to form stars, its star formation rate
can be written as
\begin{equation}
   \frac{\dd \rho_*}{\dd t} = - \frac{\dd \rho_i}{\dd t} = \frac{c_* \rho_i}{t_*}\,,
\end{equation}
where $c_*$ is a dimensionless star formation rate parameter and $t_*$
the characteristic timescale for star formation.  The value of this
timescale is usually taken to be the maximum of the dynamical time
$t_{\rm dyn}=(4\pi {\rm G} \rho_i)^{-0.5}$ and the cooling time
$t_{\rm cool}=u_i/(\dd u_i/\dd t)$.
In principle, to follow star formation, one would like to produce 
continuously collisionless star particles. However, for
computational and numerical reasons, one approximates this process by
waiting for a significant fraction of the gas particle mass to have formed stars
according to the above rate; when this is accomplished, a new, collisionless
``star'' particle is created from the parent star-forming gas particle, whose mass is
reduced accordingly. This process takes place until the gas particle
is entirely transformed into stars. In order to avoid spurious
numerical effects, which arise from the gravitational interaction of
particles with widely differing masses, one usually restricts the
number of star particles (so called {\sl generations}) spawned
by a gas particle to be relatively small, typically $2-3$. Note 
that it is also common to restrict the described star-formation algorithm
to only convert a gas particle into a star particle, 
which correspond to the choice of only one {\sl generation}. In this
case star and gas particles have always the same mass.

To get a more continuous distribution of star particle masses, the
probability of forming a star can be written as
\begin{equation}
   p = 1 - {\rm exp}\left(-c_*\frac{\Delta t}{t_g}\right)
\end{equation}
and a random number is used to decide when to form a star particle.

According to this scheme of star formation, each star particle can be
identified with a Simple Stellar Population (SSP), i.e. a coeval
population of stars characterised by a given assumed initial mass
function (IMF).  Further, assuming that all stars with masses larger
than 8~M$_\odot$ will end as type-II supernovae (SN\,II), one can
calculate the total amount of energy (typically $10^{51}$~erg per
supernova) that each star particle can release to the surrounding
gas. Under the approximation that the typical lifetime of massive
stars which explode as SN\,II does not exceed the typical time step of
the simulation, this is done in the so--called ``instantaneous
recycling approximation'', with the feedback energy deposited in
the surrounding gas in the same step.

Improvements with respect to this model include
an explicit sub--resolution description of the multi--phase nature of
the interstellar medium, which provides the reservoir of star
formation. Such a sub grid model
tries to model the global dynamical behaviour of the
interstellar medium in which cold, star-forming clouds are embedded in a
hot medium.

One example is the multi-phase sub-grid model suggested by
\citet{springel2003}, in which a star-forming gas particle has a
multi--phase nature, with a cold phase describing cold clouds embedded
in pressure equilibrium within a hot medium.
According to this model, star formation takes place in a
self--regulated way. Within this model (as within other models), part of the
feedback energy is channelled back into kinetic energy, effectively
leading to a quasi self-consistent modelling of galactic outflows,
driven by the star-forming regions.
Once an efficient form of
kinetic feedback is included, the amount of stars formed in simulations
turns out to converge when resolution is varied. An example of the predicted
stellar component in different galaxy clusters for varying spatial
resolutions, taken from \citet{2006MNRAS.367.1641B} is shown in 
Fig.~\ref{figs:coldbar}.

A further direction for improvement is provided by a more accurate
description of stellar evolution, and of the chemical enrichment
associated with star forming particles. More accurate models require
that energy feedback and metals are released not just by SN\,II, but
also by SN\,Ia and low and intermediate mass stars, thereby avoiding the
instantaneous recycling approximation (e.g., see \citealt{borgani2008b} - 
Chapter 18, this volume, for a more detailed discussion of
this point).

\subsection{Additional physics}

A number of other physical processes, besides those related to star
formation and SN feedback, are in general expected to play a role in
the evolution of the cosmic baryons and, as such, should be added into
the treatment of the hydrodynamical equations. For instance, magnetic
fields or the effect of a population of relativistic particles as
described in detail in the review by \citealt{dolag2008a} - Chapter 12, 
this volume.
Other processes, whose effect has been studied in cosmological
simulations of galaxy clusters include thermal conduction
\citep{2004MNRAS.351..423J,2004ApJ...606L..97D}, radiative transfer to describe the
propagation of photons in a medium 
(e.g. \citealt{2006MNRAS.371.1057I} and references
therein), growth of black holes
and the resulting feedback associated with the extracted energy (e.g.
\citealt{2007MNRAS.380..877S} and references therein.
A description of these processes is far outside
the scope of this review and we point the interested reader to the
original papers cited above.

\section{Connecting simulations to observations}

\begin{figure*}
\begin{center}
\includegraphics[width=0.9\textwidth]{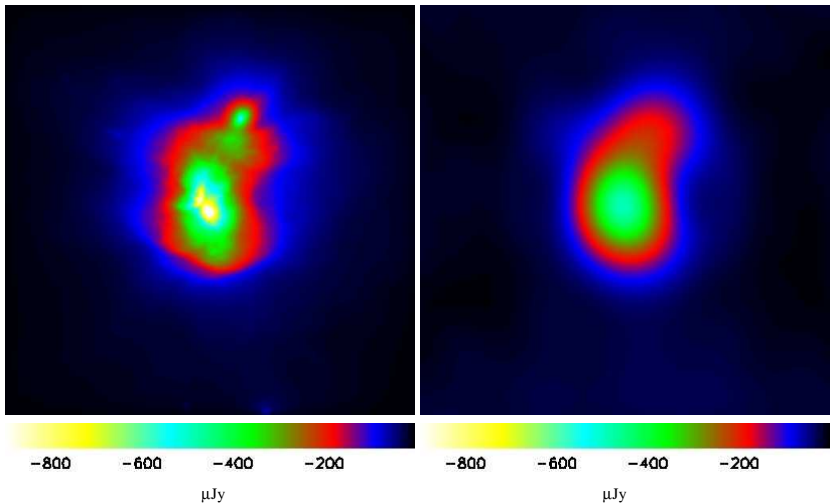}
\end{center}
\caption{Maps for the SZ decrement for a simulated galaxy cluster. The original map extracted
from the hydrodynamic simulation, and the same map in the simulated
observation ($t=34$~hour) which assumes the {\sl AMI} interferometric response, are
shown in the left and right panel, respectively. The side of each map
corresponds to 16 arcmin. The colour scale is shown at the bottom of
each panel. Taken from \citet{2007MNRAS.378.1248B}.}
\label{fig:sz}
\end{figure*}

Both the recent generation of instruments and an even more sophisticated next
generation of instruments (at various wavelengths) will allow us
to study galaxy clusters in rich detail. Therefore, when
comparing observations with simulations, instrumental effects like
resolution and noise, as well as more subtle effects within the
observational processes, have to be taken into account to separate
true features from instrumental effects and biases. Therefore, the
building up of synthetic instruments to ``observe'' simulations becomes
more and more important.

As an example, Fig.~\ref{fig:sz} shows
the difference between a Sunyaev-Zel'dovich map obtained from a simulated
galaxy cluster, and how it would be observed with the
AMI\footnote{http://www.mrao.cam.ac.uk/telescope/ami} instrument,
assuming the response according to the array configuration of the
radio dishes and adding
the appropriate noise. Here a $34$~hour observation has been assumed, and the {\sl
CLEAN} deconvolution algorithm applied.
For details see \citet{2007MNRAS.378.1248B}.

\begin{figure*}
\begin{center}
\includegraphics[width=0.6\textwidth]{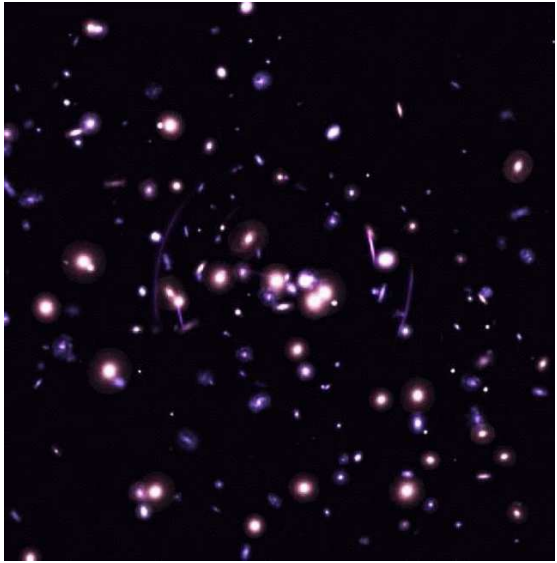}
\end{center}
\caption{Composite $ugr+riz+izy$ image of a simulated galaxy
cluster including its simulated lensing signal, having its imprint in several
strong lensing features. To construct the surface brightness distributions of
both the cluster members and of the background galaxies, shapelet
decompositions were used. For each cluster galaxy the morphological
classifications and spectral energy distribution inferred from the
semi-analytic modelling based on the merger trees from the underlying
cluster simulation were used to realistically model the optical
appearance of the cluster. The field of view is $100^{\prime\prime}\times 
100^{\prime\prime}$ and an
exposure of 1000~s for each band was assumed.
Taken from \citet{2007arXiv0711.3418M}.}
\label{fig:opt}
\end{figure*}

Another example of a synthetic observation by a virtual telescope is
shown in Fig.~\ref{fig:opt}, which shows an optical image of a
simulated galaxy cluster with several strong lensing features. Here,
\citet{2007arXiv0711.3418M} investigated the capability of the planned
{\sl DUNE} mission to measure the properties of gravitational arcs,
including instrumental effects as well as the disturbance by light from the cluster galaxies.
Shapelet decompositions - based on galaxy images retrieved
from the {\sl GOODS-ACS} archive - were used to describe the surface
brightness distributions of both the cluster members and the
background galaxies. For the cluster members the morphological
classifications inferred from the semi-analytic modelling, based on the
merger tree of the underlying cluster simulations were used to
assign a spectral energy distribution to each. Several
sources of noise like photon noise, sky background, seeing, and
instrumental noise are included. For more details see \citet{2007arXiv0711.3418M}.

Thanks to its high temperature (around $10^8$~K) and its
high density (up to $10^{-3}$ particles per cm$^3$) the intra-cluster
medium is an ideal target to be studied by X-ray telescopes.
Therefore, so far, most effort towards understanding systematic in the
observational process of galaxy clusters has been spent
interpreting X-ray observations.

For a direct comparison of the simulations with X-ray observations one
has to calculate from the simulated physical quantities (density,
temperature, $\ldots$) the observed quantities (e.g. surface
brightness). The other way -- going from observations to physical
quantities  -- is much more difficult and would require a number of
assumptions.  Fortunately, the ICM is usually optically thin, so that
absorption of photons within the ICM does not have to be taken into
account. Hence to obtain an X-ray image of the modelled cluster, one
has to choose a projection direction and integrate over all the
emission of the elements along the line of sight for each pixel in the
image. The X-ray emission at each element is usually approximated as
the product of the square of the density and the cooling function.  As
X-ray detectors are only sensitive in a certain energy range, one must
be careful to take the correct range.  For very realistic images one
needs to taken into account also the effects of the X-ray telescopes
and detectors (e.g. the limited resolution of the X-ray telescope
(point spread function) or the energy and angle dependent sensitivy
(detector response matrix and vignetting). As observed images also depend
on the distance of the cluster and the (frequency dependent)
absorption of the X-ray emission in the Galaxy these should also be
accounted for. For an exact prediction, the emitted spectrum of each element has
to be taken into account, as well as many of the effects
mentioned above are energy dependent.

Even more difficult is the comparison of quantities derived from
spectral analyses such as ICM temperature and ICM metallicity, or their
projections in profiles and maps. A pixel in an observed temperature map, for
example, is derived using all the photons within the pixel area,
and the derived spectrum is fitted with a single model of a hot plasma (see 
\citealt{kaastra2008} - Chapter 9, this volume). These photons, however, come from different
positions along the line of sight, that have different emissivities,
different temperature and different metallicities. So the spectrum is
actually a composite of many different spectra. Obviously such a
multi-temperature spectrum cannot be fitted very well by a single
temperature model.  Sometimes temperature
maps are produced in an even cruder way using the ratio of two or more
energy bands. These are called hardness ratio maps.

\begin{figure*}
\begin{center}
\includegraphics[width=0.8\textwidth]{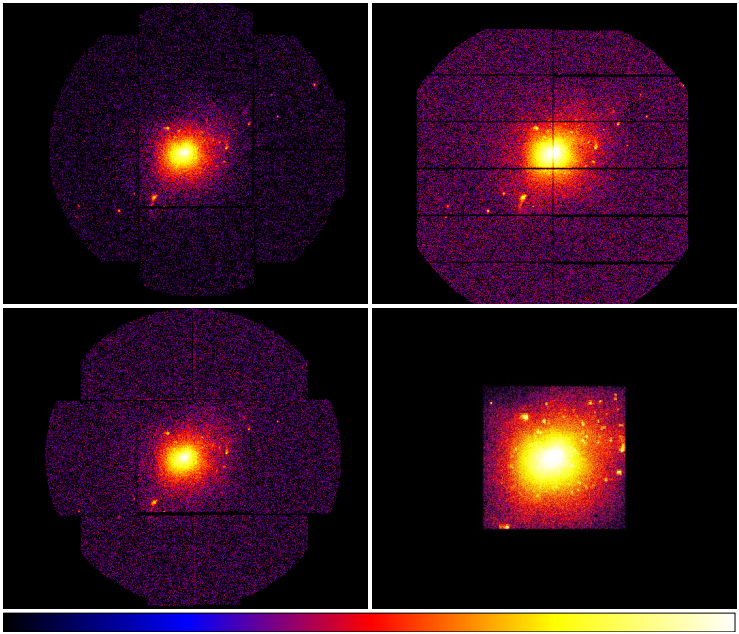}
\end{center}
\caption{Simulated photon images in the $0.7-2$~keV energy band of a
simulated galaxy cluster using {\sl XMAS-2}. The images are binned to 3.2
arcsec. They include background, vignetting effects,
out-of-time events and the telescope optical paths. From top left to
bottom right are simulations for the MOS1, PN, and MOS2 instruments on
board of the {\sl XMM-Newton} satellite and for the ACIS-S3 instrument onboard
of the {\sl Chandra} satellite. Kindly provided by Elena Rasia, see
\citet{2006MNRAS.369.2013R} and \citet{2007arXiv0707.2614R}.}
\label{fig:xmm}
\end{figure*}

To calculate temperatures, temperature profiles or temperature maps
are calculated using mostly emission weighted temperatures, i.e. summing
the temperature of all elements along the line of sight,
weighted by their emission

\begin{equation}
T = { \int W T \dd V \over \int W \dd V}
\end{equation}
with $T$ being the gas temperature and $W$ a weighting factor. Usually $W$
is proportional to the emissivity of each gas element,

\begin{equation}
W = \Lambda (T) n^2
\end{equation}
where $\Lambda (T)$ is the cooling curve and $n$ the gas density.

This simple procedure does not of course
take into account the shape of the spectra corresponding to gas of
different temperatures. It therefore gives only a rough estimate of
the actual temperature. With numerical simulations it was investigated how
accurate these emission weighted temperatures are, by comparing them
with temperatures obtained by actually adding spectra, so-called
spectroscopic temperatures. It was found that the emission-weighted
temperatures are systematically higher than the spectroscopic
temperatures
\citep{2001ApJ...546..100M,2004MNRAS.351..505G,2004MNRAS.354...10M}.
To overcome this problem, \citet{2004MNRAS.354...10M} suggested
an approximation to the spectroscopic temperature,
the ``spectroscopic-like'' temperature $T_{\rm sl}$

\begin{equation}
T_{\rm sl} = { \int n^2 T^\alpha/T^{1/2} \dd V
\over    \int n^2 T^\alpha/T^{3/2} \dd V    }
\end{equation}
which yields for $\alpha = 0.75$ a good estimate of the  spectroscopic
temperature. In addition the simulated $M$-$T$ relation is strongly
affected, if the emission-weighted temperature is used \citep{2005ApJ...618L...1R}.

The inhomogenous temperature and metal distribution was also found to be
responsible also for inaccurate metallicity measurements. \citet{2007arXiv0707.2614R}
studied with numerical simulations, together with the
programme {\sl X-MAS2}, how well the elements Fe, O, Mg and Si can be
measured in clusters of different temperature. They find that Fe and Si are
generally measureable with good accuracy, while O and Mg can be considerably
overestimated in hot clusters. Using simulations and the programme
{\sl SPEX} (see
\citealt{kaastra2008} - Chapter 9 this volume)
\citet{2007A&A...472..757K} found that due to the metal
inhomogeneities the metal mass in clusters is systematically
underestimated -- in extreme cases by up to a factor of three.

An example of synthetic X-ray observations is shown in Fig.~\ref{fig:xmm}. 
Here {\sl  XMAS-2} is used to produce photon images
from a simulated galaxy cluster. Shown are simulations for the MOS1, PN,
and MOS2 instruments on board the {\sl XMM-Newton} satellite, and for the
ACIS-S3 instrument on board the {\sl Chandra} satellite. They include
instrumental effects like background, vignetting and response for the
individual instruments. For more details see \citet{2006MNRAS.369.2013R}
and \citet{2007arXiv0707.2614R}.

\section{Outlook}

In the future, the demand on precision in both simulation techniques
and captured complexity of the physical processes within the simulations will be
quite challenging. Recent leading simulations are already extremely
difficult to analyse due to their enormous size and complexity, and they
will surely continue to grow. In fact, the next generation of
supercomputers will grow more in the number of accessable CPUs than on the
speedup of the individual CPUs and this fact will make the analysis of the
simulations as challenging as performing the simulations itself.
To keep a comparable level of accuracy, the interpretation of a simulation
of the next generation of high precision experiments will need
to massively involve virtual telescopes as described in the previous
section. This will increase the need of involving complex analysis
pipelines for ``observing'' simulations, and might lead to a new branch of
{\sl virtual observers} in the astrophysics community, similar to the
already, new formed branch of {\sl computational astrophysicists}.

\section{Acknowledgments}
The authors thank ISSI (Bern) for support of the team
``Non-virialized X-ray components in clusters of galaxies''.
Special thanks to Volker Springel for helping to improve the
manuscript and providing Fig.~\ref{figs:moor_nbody}
to Anna Watts for carefully reading the manuscript,
to Martin Obergaulinger for very helpful discussions on Eulerian
schemes, to Ewald M\"uller for the lecture notes on
the Riemann problem and providing Fig.~\ref{figs:riemann}, 
to Elena Rasia for providing
Fig.~\ref{fig:xmm} and to Umberto Maio for providing Fig.~\ref{figs:cooling}.
A.D. also gratefully acknowledges
partial support from the PRIN2006 grant ``Costituenti fondamentali
dell'Universo'' of the Italian Ministry of University and
Scientific Research and from the INFN grant PD51.

\bibliographystyle{aa}
\bibliography{12_dolag}

\end{document}